\documentclass[aps,prd,twocolumn,a4paper,groupedaddress]{revtex4-1}
\pdfoutput=1 
\usepackage{atlasphysics} 
\usepackage{upgreek}
 \usepackage[hyperindex,breaklinks]{hyperref} 
\usepackage{rotating}
\usepackage{amsmath}

\newcommand{\ilumi} {$ 4.9\, \rm fb^{-1} $}
\newcommand{\nsig} {\ensuremath{ 22670 \pm 150 }}
\newcommand{\DGfit}     { 0.053 }
\newcommand{\DGstat}    { 0.021 }
\newcommand{\DGsyst}    { 0.010 }
\newcommand{\pSfit}     { 0.12 }
\newcommand{\pSstat}    { 0.25 }
\newcommand{\pSsyst}    { 0.05 }
\newcommand{\GSfit}     { 0.677 }
\newcommand{\GSstat}    { 0.007 }
\newcommand{\GSsyst}    { 0.004 }
\newcommand{\Azesqfit}     { 0.529 }
\newcommand{\Azesqstat}    { 0.006 }
\newcommand{\Azesqsyst}    { 0.012 }
\newcommand{\Apasqfit}     { 0.220 }
\newcommand{\Apasqstat}    { 0.008 }
\newcommand{\Apasqsyst}    { 0.009 }
\newcommand{\ASsqfit}      { 0.024 }
\newcommand{\ASsqstat}     { 0.014 }
\newcommand{\ASsqsyst}     { 0.028 }
\newcommand{\Deltperpfit}      { 3.89 }
\newcommand{\Deltperpstat}     { 0.47 }
\newcommand{\Deltperpsyst}     { 0.11 }
\newcommand{\Deltparafit}     { [3.04, 3.23] }
\newcommand{\Deltparasyst}     { 0.09 }
\newcommand{\DeltPSfit}        { [3.02, 3.25]}
\newcommand{\DeltPSsyst}        { 0.04}
\newcommand{\pAz}   { 0.002 }  
\newcommand{\pApa}  { 0.010 }   
\newcommand{\pAs}   {  0.029 }  
\newcommand{\pG}    {  0.026 } 
\newcommand{\pDG}   { 0.107 }   
 \newcommand{\pDpara}   { 0.021 }
\newcommand{\pDperp}   {  -0.043 }
 \newcommand{\pDpS}   {  -0.003 }  
\newcommand{\DGAz}  {  0.103 } 
\newcommand{\DGApa} {  0.105 }   
\newcommand{\DGAs}  {  0.069 }  
\newcommand{\DGG}   { -0.617 } 
 \newcommand{\DGDpara}   { 0.006 }
 \newcommand{\DGDperp}   {  -0.017 }
 \newcommand{\DGDpS}   {  0.001 } 
\newcommand{\GAz}   { -0.063 }  
\newcommand{\GApa}  { -0.093 }  
\newcommand{\GAs}   {  0.034 }  
 \newcommand{\GDpara}   { -0.003  }
 \newcommand{\GDperp}   {  0.001 }
 \newcommand{\GDpS}   {  -0.009 } 
\newcommand{\AzApa} { -0.316 } 
\newcommand{\AzAs}  {  0.283 }  
 \newcommand{\AzDpara}   { - 0.003 }
 \newcommand{\AzDperp}   {  -0.016   }
 \newcommand{\AzDpS}   {  -0.025 } 
\newcommand{\ApaAs} {  0.077 }  
\newcommand{\ApaDpara}   { 0.008 }
 \newcommand{\ApaDperp}   {  0.005 }
 \newcommand{\ApaDpS}   {  -0.010 } 
\newcommand{\AsDpara}   { -0.011 }
 \newcommand{\AsDperp}   {   -0.054 }
 \newcommand{\AsDpS}   {  -0.098 } 
\newcommand{\DparaDperp}   {  0.038 }
 \newcommand{\DparaDpS}   {  0.007 } 
\newcommand{\DperpDpS}   {  0.081 } 
\newcommand{\cutMassJpsiBB} {$ (2.959$--$3.229) ~\rm GeV $}
\newcommand{\cutMassJpsiEB} {$ (2.913$--$3.273) ~\rm GeV $}
\newcommand{\cutMassJpsiEE} {$ (2.852$--$3.332) ~\rm GeV $}
\newcommand{\cutChiJpsi}   {\chisq $< 10$}
\newcommand{\cutMassPhi} {$ 1.0085 \; {\rm GeV} < m(K^+ K^-) < 1.0305 \; {\rm GeV} $}
\newcommand{\cutChiBs}   {\chisq $< 3$}
\newcommand{\Bst}{\ensuremath{ \Bs \ra \JpsiPhi }} 
\newcommand{\Bsto}{\ensuremath{ \Bs \ra J/\psi (\mumu) \phi(K^+K^-) }} 
\newcommand{\BstoKK}{\ensuremath{ \Bs \ra J/\psi K^+K^- (f_{0}) }} 
\newcommand{\JpsiPhi}{\ensuremath{ J/\psi\phi }} 
\newcommand{\Gs}{\ensuremath{ \Gamma_{s}     }}
\newcommand{\DGs}{\ensuremath{ \Delta\Gamma_{s}     }} 
\newcommand{\phis}{\ensuremath{\phi_{s}  }} 
\newcommand{\dms}{\ensuremath{\Delta m_{s}  }} 
\newcommand{\BH}{\ensuremath{B_{H}  }} 
\newcommand{\BL}{\ensuremath{B_{L}  }} 
\newcommand{\GL}{\ensuremath{\Gamma_{L}  }}

\newcommand{\BsaBs}{\ensuremath{\Bs - \aBs}} 
\newcommand{\B}{\ensuremath{ B}\space  }

\newcommand{\aBs}{\ensuremath{\overline{B}^0_s}} 

\newcommand{\Jpsipa}{\Jpsi \ }

\newcommand{\chisq}{$\rm \chi^{2}$/d.o.f. }
\newcommand{\Jpsitomm}{\ensuremath{ \Jpsi \ra \mumu }\space} 
\newcommand{\phiKK}{\ensuremath{ \phi \ra \kplus \kminus}\space}

\usepackage{preprintcover}  

\PreprintCoverPaperTitle{Flavour tagged time dependent angular analysis of the $\boldsymbol{\Bst}$ decay and extraction of $\boldsymbol{\DGs}$ and the weak phase $\boldsymbol{\phis}$ in ATLAS}  

\PreprintIdNumber{CERN-PH-EP-2014-043}  

\PreprintCoverAbstract{A measurement of the \Bst\ decay parameters, updated to include flavour tagging is reported using \ilumi\ of integrated luminosity collected by the ATLAS detector from $\rts = 7\TeV$ $pp$ collisions recorded in 2011 at the LHC. The values measured for the physical parameters are:
\begin{eqnarray*}
\phis & = & \pSfit \pm \pSstat~\rm{(stat.)}\pm \pSsyst~\rm{(syst.)}~rad  \\
\DGs & = & \DGfit \pm \DGstat~\rm{(stat.)}\pm \DGsyst~\rm{(syst.)~ps}^{-1} \\
\Gs & = & \GSfit \pm \GSstat ~\rm{(stat.)}\pm \GSsyst ~\rm{(syst.)~ps}^{-1}\\
|A_{\parallel}(0)|^2 & = & \Apasqfit \pm \Apasqstat~\rm{(stat.)}\pm \Apasqsyst~\rm{(syst.)}\\
|A_{0}(0)|^2 & = & \Azesqfit \pm \Azesqstat~\rm{(stat.)}\pm \Azesqsyst~\rm{(syst.)}\\
\delta_\perp & = & \Deltperpfit \pm \Deltperpstat ~\rm{(stat.)}\pm \ \Deltperpsyst~\rm{(syst.)}~rad
\end{eqnarray*}
where the parameter \DGs ~ is constrained to be positive.  The $S$-wave contribution was measured and found to be compatible with zero.  Results for \phis\ and \DGs\ are also presented as 68\% and 95\% likelihood contours, which show agreement with the Standard Model expectations.}  

\PreprintJournalName{Physical Review D}

\begin{document}

\title{Flavour tagged time dependent angular analysis of the $\boldsymbol{\Bst}$ decay and extraction of $\boldsymbol{\DGs}$ and the weak phase $\boldsymbol{\phis}$ in ATLAS}

\author{The ATLAS Collaboration}
\noaffiliation

\date{\today}

\begin{abstract}
A measurement of the \Bst\ decay parameters, updated to include flavour tagging is reported using \ilumi\ of integrated luminosity collected by the ATLAS detector from $\rts = 7\TeV$ $pp$ collisions recorded in 2011 at the LHC. The values measured for the physical parameters are:
\begin{eqnarray*}
\phis & = & \pSfit \pm \pSstat~\rm{(stat.)}\pm \pSsyst~\rm{(syst.)}~rad  \\
\DGs & = & \DGfit \pm \DGstat~\rm{(stat.)}\pm \DGsyst~\rm{(syst.)~ps}^{-1} \\
\Gs & = & \GSfit \pm \GSstat ~\rm{(stat.)}\pm \GSsyst ~\rm{(syst.)~ps}^{-1}\\
|A_{\parallel}(0)|^2 & = & \Apasqfit \pm \Apasqstat~\rm{(stat.)}\pm \Apasqsyst~\rm{(syst.)}\\
|A_{0}(0)|^2 & = & \Azesqfit \pm \Azesqstat~\rm{(stat.)}\pm \Azesqsyst~\rm{(syst.)}\\
\delta_\perp & = & \Deltperpfit \pm \Deltperpstat ~\rm{(stat.)}\pm \ \Deltperpsyst~\rm{(syst.)}~rad
\end{eqnarray*}
where the parameter \DGs ~ is constrained to be positive.  The $S$-wave contribution was measured and found to be compatible with zero.  Results for \phis\ and \DGs\ are also presented as 68\% and 95\% likelihood contours, which show agreement with the Standard Model expectations. 
\end{abstract}


\maketitle

\section{Introduction}
New phenomena beyond the predictions of the Standard Model (SM) may alter $CP$ violation in $B$-decays. A channel that  is expected to be sensitive to  new physics contributions is the decay \Bst. $CP$ violation in the \Bst\ decay occurs due to interference between direct decays and decays with \BsaBs\ mixing.  The oscillation frequency of \Bs\ meson mixing is characterized by the mass difference \dms\ of the heavy (\BH) and light (\BL)  mass eigenstates. The $CP$ violating phase \phis\ is defined as the weak phase difference between the \BsaBs\ mixing amplitude and the $b \rightarrow c \overline{c} s$ decay amplitude.  In the absence of $CP$ violation, the \BH\ state would correspond to the $CP$ odd state and the \BL\ to the $CP$ even state.  In the SM the phase \phis\ is small and can be related to CKM quark mixing matrix elements via the relation $\phis \simeq -2 \beta_{s}$, with $\beta_{s} = \mathrm{arg} [- (V_{ts} V^{*}_{tb})/(V_{cs} V_{cb}^{*}) ]$; a value of  $\phis \simeq -2 \beta_{s} = -0.037 \pm 0.002 $ rad \cite{PHISM} is predicted in the SM. Many new physics models predict large  \phis\ values whilst satisfying all existing constraints, including the precisely measured value of  \dms\  \cite{dmsTevatron,LHCboscil}.

Another physical quantity involved in \BsaBs\ mixing is the width difference $\DGs = \GL - \GH$, which is predicted to be $\DGs = 0.087 \pm 0.021$ ps$^{-1}$ \cite{DeltaGamma}.  Physics beyond the SM is not expected to affect \DGs\ as significantly as \phis~\cite{Nierste}. Extracting \DGs\ from data is nevertheless useful as it allows theoretical predictions to be tested~\cite{Nierste}.

The decay of the pseudoscalar \Bs\ to the vector--vector  final-state $J/\psi\phi$  results in an admixture of $CP$ odd and $CP$ even states, with orbital angular momentum $L = 0$, 1 or 2. The final states with orbital angular momentum $L = 0$ or 2 are $CP$ even while the state with $L = 1$ is $CP$ odd.  Flavour tagging is used to distinguish between the initial \Bs\ and \aBs\ states.  The $CP$ states are separated statistically using an angular analysis of the final-state particles. 

In this paper, an update to the previous measurement \cite{ATLASBsJpsiphi} with the addition of flavour tagging, is presented.  Flavour tagging significantly reduces the uncertainty of the measured value of \phis\ while also allowing a measurement of one of the strong phases.  Previous measurements of these quantities have been reported by the D\O , CDF and LHCb collaborations \cite{Abazov:2011ry, CDF:2012ie, LHCb:2013oba}. The analysis presented here uses \ilumi\ of LHC {\it pp} data at $\sqrt{s} = 7 \TeV$ collected by the ATLAS detector in 2011.

\section{ATLAS detector and Monte Carlo simulation} \label{sec:sample}
The ATLAS experiment \cite{ATLAS} is a multipurpose particle physics detector with a forward-backward symmetric cylindrical geometry and near $4\pi$ coverage in solid angle. The inner tracking detector (ID) consists of a silicon pixel detector, a silicon microstrip detector  and a transition radiation tracker. The ID is surrounded by a thin superconducting solenoid providing a 2\,T axial magnetic field, and by a high granularity liquid-argon sampling electromagnetic calorimeter. A steel/scintillator tile calorimeter provides hadronic coverage in the central rapidity range. The end-cap and forward regions are instrumented with Liquid-argon calorimeters for both electromagnetic and hadronic measurements. The muon spectrometer (MS) surrounds the calorimeters and consists of three large superconducting toroids with eight coils each, a system of tracking chambers, and detectors for triggering.

The muon and tracking systems are of particular importance in the reconstruction of \B meson candidates. Only data where both systems were operating correctly and where the LHC beams were declared to be stable are used. A muon identified using a combination of MS and ID track parameters is referred to as {\it combined}. A muon formed by track segments which are not associated with an MS track, but which are matched to ID tracks extrapolated to the MS is referred to as {\it  segment tagged}. 

The data were collected during a period of rising instantaneous luminosity, and the trigger conditions varied over this time.  The triggers used to select events for this analysis are based on identification of a \Jpsitomm decay, with either a $4\;\mathrm{GeV}$ transverse momentum \footnote{ATLAS uses a right-handed coordinate system with its origin at the nominal interaction point (IP) in the centre of the detector and the $z$-axis along the beam pipe. The $x$-axis points from the IP to the centre of the LHC ring, and the $y$-axis points upward. Cylindrical coordinates $(r,\phi)$ are used in the transverse plane, $\phi$ being the azimuthal angle around the beam pipe. The pseudorapidity is defined in terms of the polar angle $\theta$ as $\eta=-\ln\tan(\theta/2)$.} (\pT) threshold for each muon or an asymmetric configuration that applies a \pT\ threshold of $4\GeV$ to one of the muons while accepting a second muon with \pT\ as low as $2\GeV$.

Monte Carlo (MC) simulation is used to study the detector response, estimate backgrounds and model systematic effects. For this study, 12 million MC-simulated \Bst\ events were generated using  {\sc Pythia} 6 \cite{Pythia} tuned with recent ATLAS data~\cite{ATL-PHYS-PUB-2011-009}. No \pT\ cuts were applied at the generator level.  Detector responses for these events were simulated using the ATLAS simulation package based on GEANT4~\cite{Atlfast, Geant4}.  Pile-up corresponding to the conditions during data taking was included.  In order to take into account the varying trigger configurations during data-taking, the MC events were weighted to have the same trigger composition as the collected collision data.  Additional samples of the  background decay $\Bd \ra J/\psi K^{0*}$ as well as the more general $bb \ra J/\psi X$ and $pp \ra J/\psi X$ backgrounds were also simulated using  {\sc Pythia}.

\section{Reconstruction and candidate selection}
Events passing the trigger and the data quality selections described in section~\ref{sec:sample} are required to pass the following additional criteria: the event must contain at least one reconstructed primary vertex, built from at least four ID tracks, and at least one pair of oppositely charged muon candidates that are reconstructed using information from the MS and the ID \cite{Aad20119}.   Both combined and segment tagged muons are used.  In this analysis the muon track parameters are taken from the ID measurement alone, since the precision of  the measured track parameters for muons in the  \pT\ range of interest for this analysis is dominated by the ID track reconstruction.  The pairs of muon tracks are refitted to a common vertex and accepted for further consideration if the fit results in \cutChiJpsi. The invariant mass of the muon pair is calculated from the refitted track parameters.  To account for  varying  mass resolution, the \Jpsi\ candidates are divided into three subsets  according to the pseudorapidity $\eta$ of the muons. A maximum likelihood fit is used to extract the $J/\psi$ mass and the corresponding resolution for these three subsets.  When both muons have $|\eta| <  1.05$, the dimuon invariant mass must fall in the range \cutMassJpsiBB \ to be accepted as a  \Jpsi\ candidate. When one muon has $1.05  < |\eta| <  2.5 $ and the other muon  $|\eta| <  1.05$, the corresponding signal region is \cutMassJpsiEB. For the third subset, where both muons have $1.05  < |\eta| <  2.5 $, the signal region is \cutMassJpsiEE .  In each case the signal region is defined so as to  retain 99.8\% of the \Jpsi\ candidates identified in the fits.  

The candidates for  \phiKK are reconstructed from all pairs of oppositely charged particles with $\pT > 0.5$~\GeV \ and $|\eta| < 2.5$ that are not identified as muons. Candidates  for  \Bsto\ are sought by fitting the tracks for each combination of \Jpsitomm\ and \phiKK to a common vertex.  Each of the four tracks is required to have at least one hit in the pixel detector and at least four hits in the silicon microstrip detector. The fit is further constrained by fixing the invariant mass calculated from the two muon tracks to the \Jpsipa mass~\cite{PDG}. These quadruplets of tracks are accepted for further analysis if the vertex fit has a  \cutChiBs, the fitted  \pT\ of each track from \phiKK\ is greater than 1 GeV and the invariant mass of the track pairs (under the assumption that they are kaons) falls within the interval \cutMassPhi. If there is more than one accepted candidate in the event, the candidate with the lowest $\chi^2 /$d.o.f. is selected.  In total $131513$ \Bs\ candidates are collected within a mass range of $5.15 < m(\Bs) < 5.65 \GeV$.

For each \Bs\ meson candidate the proper decay time $t$ is estimated by the expression:
\begin{equation*}
  t = \frac{ L_{xy}\ M_{{}_{B}} }     { p_{ \mathrm{T}_{B}  }}, 
\end{equation*}
where $p_{ \mathrm{T}_{B}  }$ is the  reconstructed transverse momentum of the \Bs\ meson candidate and  $M_{{}_{B}}$ denotes  the world average mass value \cite{PDG} of the \Bs\ meson. The transverse decay length, $L_{xy}$, is the displacement in the transverse plane of the  \Bs\ meson  decay vertex with respect to the primary vertex, projected onto the direction of the \Bs\ transverse momentum. The position of the primary vertex used to calculate this quantity is refitted following the removal of the tracks used to reconstruct the \Bs\ meson candidate.

For the selected events the average number of pileup interactions is 5.6, necessitating  a choice of the best candidate for the primary vertex at which the \Bs\  meson is  produced.  The variable used is the three-dimensional impact parameter $d_0$, which is  calculated as the distance between the line extrapolated from the reconstructed \Bs\ meson vertex in the direction of the \Bs\ momentum, and  each primary vertex candidate. The chosen primary vertex is the one with the smallest $d_0$. Using MC simulation it is shown that the fraction of \Bs \ candidates which are assigned the wrong primary vertex is less than 1\% and that the corresponding effect on the final results is negligible. No \Bs\ meson decay time cut is applied in the analysis. 

\section{Flavour tagging}\label{sec:Tagging}
The determination of the initial flavour of neutral $B$-mesons can be inferred using information from the $B$-meson  that is typically produced from the other $b$-quark in the event~\cite{Field:1977fa}. This is referred to as the opposite-side tagging (OST). 

To study and calibrate the OST methods, events containing the decays of $B^{\pm}\to J/\psi K^{\pm}$ can be used, where flavour of the $B$-meson at production is provided by the kaon charge. Events from the entire 2011 run period satisfying the same data quality selections as described in section 2 are used.

\subsection{$B^{\pm}\to J/\psi K^{\pm}$ event selection}
To be selected for use in the calibration analysis, events must satisfy a trigger condition requiring  two oppositely charged muons within an invariant mass range around the nominal $J/\psi$ mass.  Candidate $B^{\pm}\to J/\psi K^{\pm}$ decays are identified using two oppositely charged combined muons forming a good vertex using information supplied by the inner detector. Each muon is required to have a transverse momentum of at least $4$~\GeV\ and pseudo-rapidity within $|\eta|<2.5$. The invariant mass of the dimuon candidate is required to satisfy $2.8 < m(\mu^+\mu^-) < 3.4$~\GeV. To form the $B$ candidate an additional track with the charged kaon mass hypothesis, $\pt>1\GeV$ and $|\eta|<2.5$ is combined with the dimuon candidate and a vertex fit is performed with the mass of the di-muon pair constrained to the known value of the $J/\psi$ mass. To reduce the prompt component of the combinatorial background, the requirement $L_{xy}>0.1$~mm is applied to the $B$ candidate.  The choice of primary vertex is determined using the same procedure as done for the \Bs\ candidates.

In order to study the distributions corresponding to the signal processes with the background component removed, a sideband subtraction method is defined. Events are separated into five equal regions of $B$ candidate rapidity from $0 - 2.5$ and three mass regions.  The mass regions are defined as a signal region around the fitted peak signal mass position $\mu \pm 2\sigma$ and the sidebands are $[\mu - 5\sigma,\mu-3\sigma]$ and $[\mu+3\sigma, \mu+5\sigma]$, where $\mu$ and $\sigma$ are the mean and width of the Gaussian function describing the $B$ signal mass, for each rapidity region.  Individual binned extended maximum likelihood fits to the invariant mass distribution are performed in each region of rapidity.

\begin{figure}[t]
\begin{center}
\includegraphics[width=0.45\textwidth]{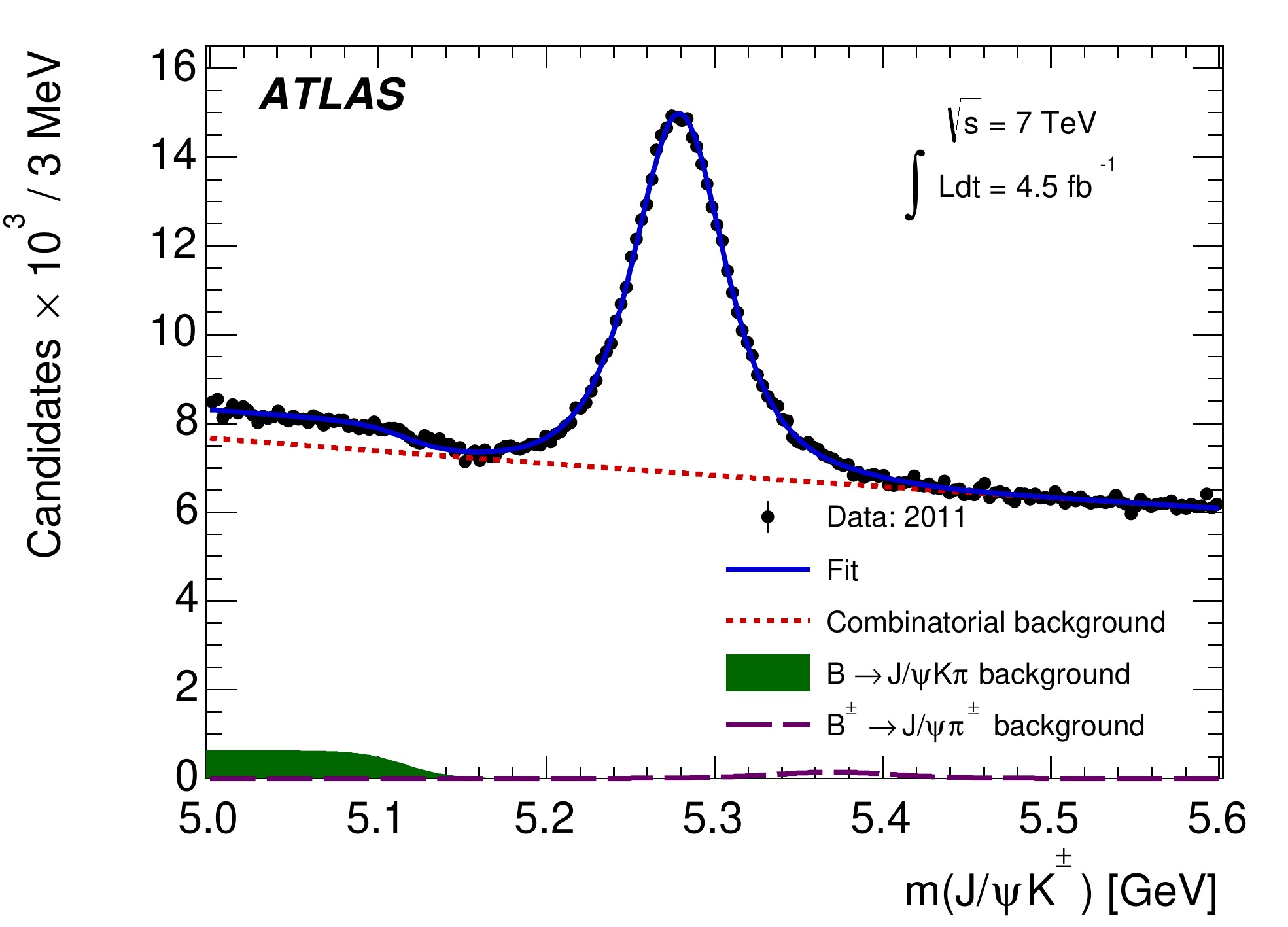}
\caption{The invariant mass distribution for  $B^{\pm}\to\jpsi K^{\pm}$ candidates. Included in this plot are all events passing the selection criteria. The data are shown by points, and the overall result of the fit is given by the blue curve.  The combinatorial background component is given by the red dotted line, partially reconstructed $B$ decays by the green shaded area, and decays of $B^{\pm}\to J/\psi \pi^{\pm}$, where the pion is mis-assigned a kaon mass by a purple dashed line.
\label{fig:bplusmass}}
\end{center}
\end{figure}

The background is modelled by an exponential to describe combinatorial background and a hyperbolic tangent function to parameterize the low-mass contribution from incorrectly or partially reconstructed $B$ decays. A Gaussian function is used to model the $B^{\pm}\to J/\psi \pi^{\pm}$ contribution. The contributions of non-combinatorial backgrounds are found to have a negligible effect in the tagging procedure. Figure~\ref{fig:bplusmass} shows the invariant mass distribution of $B$ candidates for all rapidity regions overlaid with the fit result for the combined data.

\subsection{Tagging methods}
Several methods are available to infer the flavour of the opposite-side $b$-quark, with varying efficiencies and discriminating powers. The measured charge of a muon from the semileptonic decay of the $B$ meson provides strong separation power; however, the $b\to\mu$ transitions are diluted through neutral $B$ meson oscillations, as well as by cascade decays  $b\to c\to\mu$ which can alter the sign of the muon relative to the one from direct semileptonic decays $b \to\mu$.  The separation power of tagging muons can be enhanced by considering a weighted sum of the charge of the tracks in a cone around the muon.  If no muon is present, a weighted sum of the charge of tracks associated with the opposite-side $B$ meson decay will provide some separation. The tagging methods are described in detail below.

For muon-based tagging, an additional muon is required in the event, with $\pT>2.5\GeV$, $|\eta|<2.5$ and with $|\Delta z|<5$~mm from the primary vertex. Muons are classified according to their reconstruction class, {\it combined} or {\it segment tagged}, and subsequently treated as distinct tagging methods. In the case of multiple muons, the muon with highest transverse momentum is selected.

\begin{figure}
\begin{center}
\includegraphics[width=0.49\textwidth]{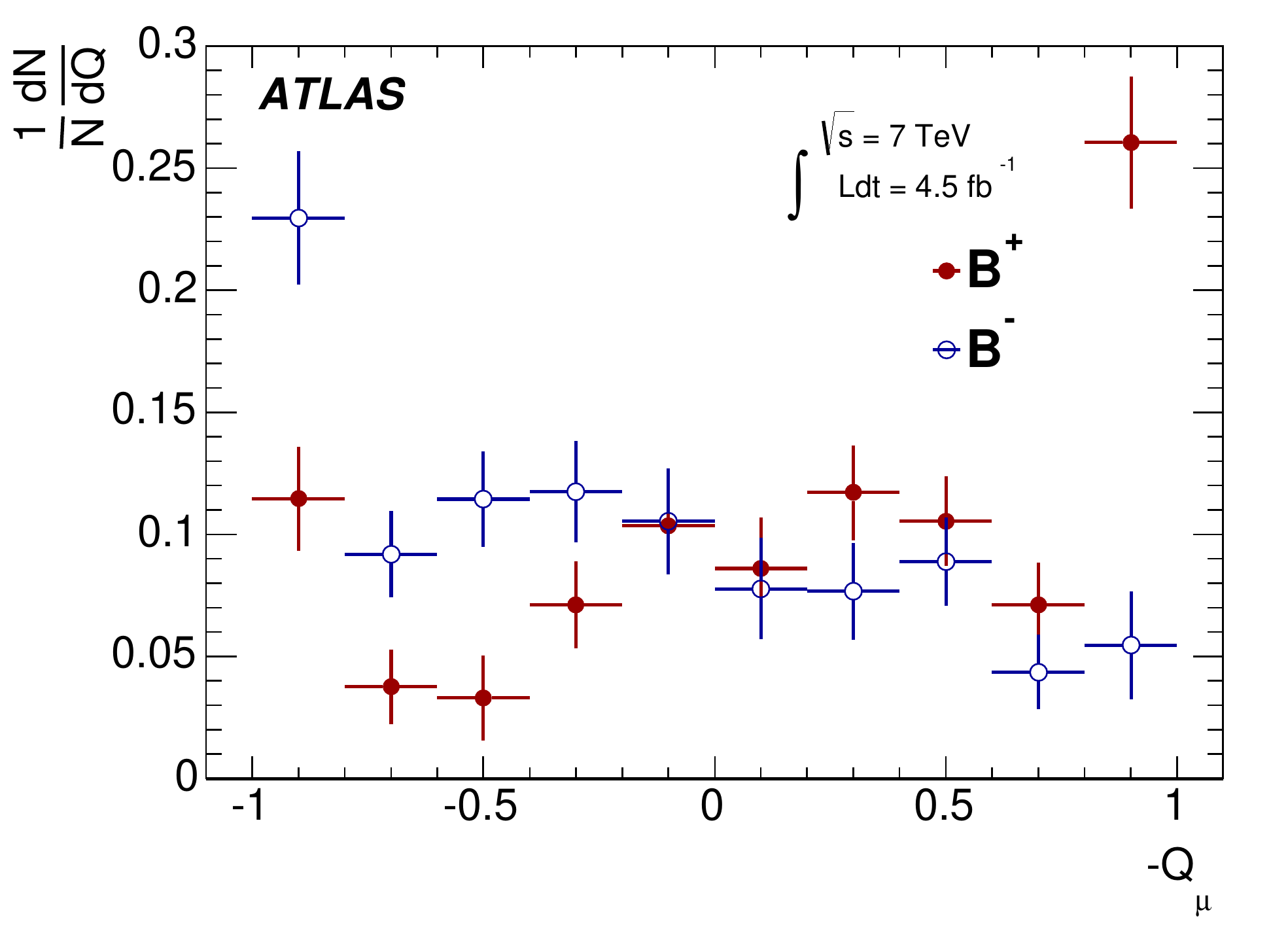}
\includegraphics[width=0.49\textwidth]{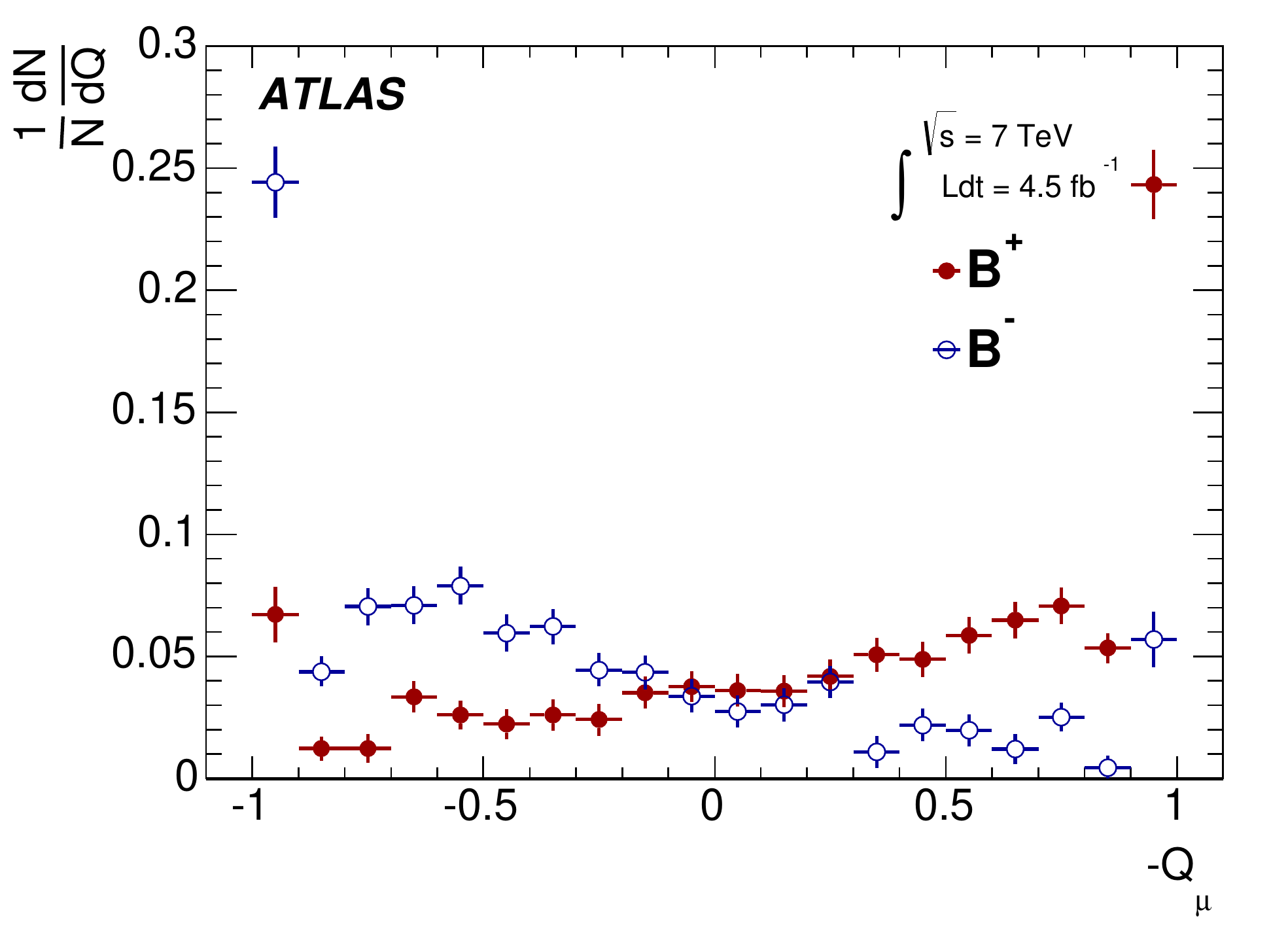}
\caption{The opposite-side muon cone charge distribution for $B^{\pm}$ signal candidates for  {\it segment tagged} (top) and {\it combined} (bottom) muons.
\label{fig:tagmudist}}
\end{center}
\end{figure}

A muon {\it cone charge} variable is constructed, defined as
\begin{equation*}
Q_{\mu}  =  \frac{ \sum^{N\;{\rm tracks}}_{i} q^{i} \cdot (\pt^{i})^{\kappa}} {  \sum^{N\;{\rm tracks}}_{i}(\pt^{i})^{\kappa}},
\end{equation*} 
where $q$ is the charge of the track, $\kappa=1.1$ and the sum is performed over the reconstructed ID tracks within a cone size of $\Delta R = 0.5$ \footnote{$\Delta R^2 = \Delta\phi^2 + \Delta\eta^2$, where $\Delta\phi$ and $\Delta\eta$ are the differences between the measured $\phi$ and $\eta$ of the tracks respectively.} around the muon direction and the muon track is included as well.   The reconstructed ID tracks must have a $\pt>0.5$~GeV and $|\eta|<2.5$. The value of the parameter $\kappa$ was determined while optimizing the tagging performance. Tracks associated with the signal decay are explicitly excluded from the sum. In figure~\ref{fig:tagmudist} the opposite-side muon cone charge distributions are shown for candidates from $B^{\pm}$ signal decays.
In the absence of a muon, a b-tagged jet \cite{ATLAS-CONF-2011-102} is required in the event, which is seeded from calorimeter clusters, with minimum energy threshold of $10~\GeV$, and where a minimum b-tag weight requirement of at least $-0.5$ is applied.  The jet tracks are required to be associated with the same primary vertex as the signal decay, excluding those from the signal candidate. Jets within a cone of $\Delta R < 0.5$ of the signal momentum axis are excluded.  The jet is reconstructed using the anti-${k_t}$ algorithm with a cone size of $0.6$. In the case of multiple jets, the jet with the highest value of the $b$-tag weight is used. 

\begin{figure}[b]
\begin{center}
\includegraphics[width=0.49\textwidth]{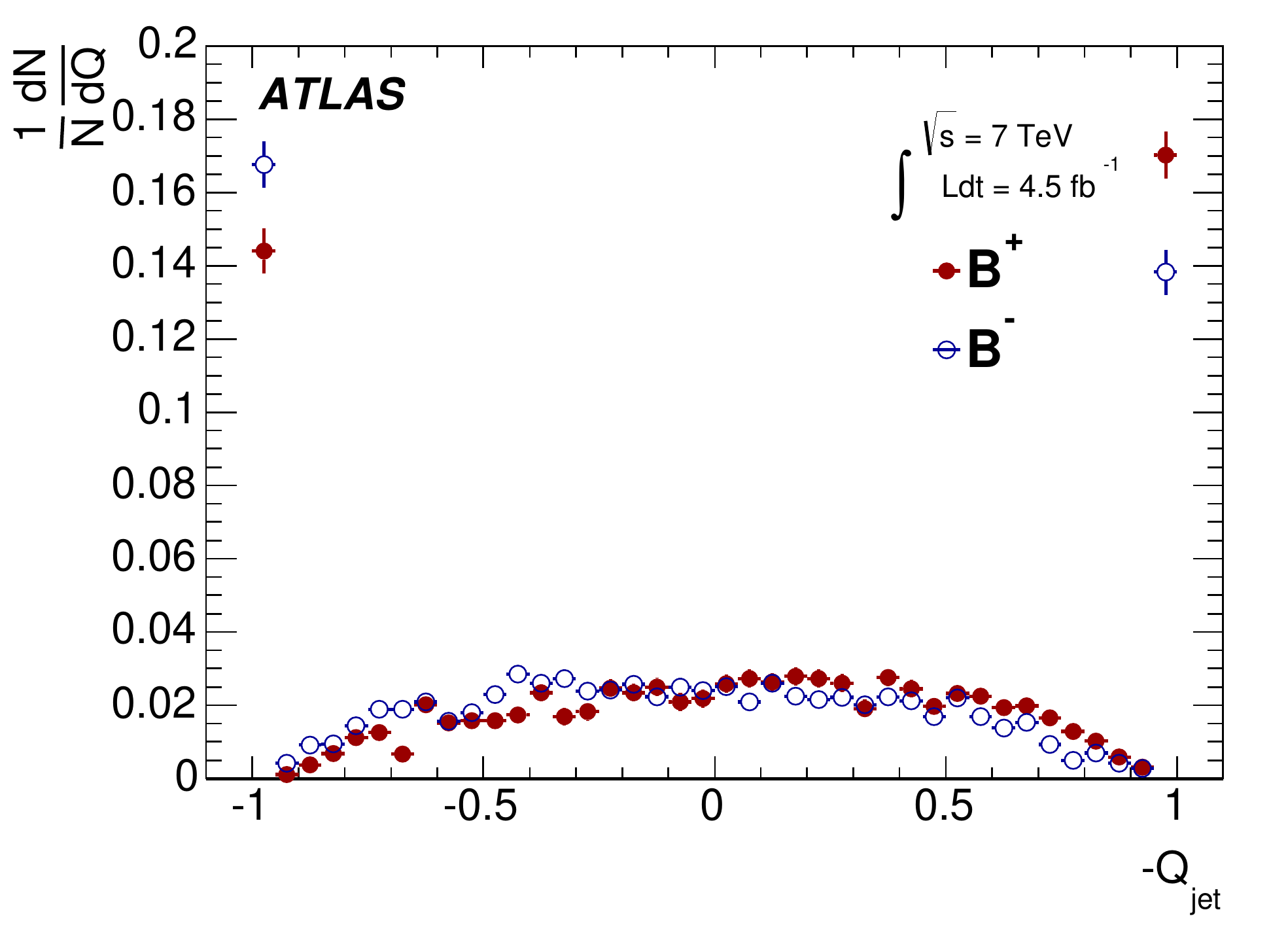}
\caption{Jet-charge distribution for $B^{\pm}$ signal candidates.
\label{fig:jetchargedist}}
\end{center}
\end{figure}

A {\it jet charge} is defined as
\begin{equation*}
Q_{\rm jet}  =  \frac{ \sum^{N\;{\rm tracks}}_{i} q^{i} \cdot (p_{T}^{i})^{\kappa}} {  \sum^{N\;{\rm tracks}}_{i}(p_{T}^{i})^{\kappa}},
\end{equation*}
where $\kappa=1.1$, and the sum is over the tracks associated with the jet, using the method described in ref. \cite{Aad:1125884}. Figure~\ref{fig:jetchargedist} shows the distribution of charges for opposite-side jet charge from  $B^{\pm}$ signal candidate events.

\begin{table*}
\begin{center}
\begin{ruledtabular}
\begin{tabular}{l|c|c|c}
Tagger & Efficiency $[\%]$ & Dilution $[\%]$ & Tagging Power $[\%]$ \\
\hline\hline
Combined $\mu$   & $ 3.37   \pm 0.04 $ & $ 50.6  \pm 0.5  $   & $ 0.86 \pm 0.04$ \\
Segment Tagged $\mu$ & $ 1.08   \pm 0.02 $ & $ 36.7  \pm 0.7 $    & $ 0.15 \pm 0.02$ \\
Jet charge            &  $ 27.7   \pm 0.1 $ & $ 12.68  \pm 0.06 $  & $ 0.45 \pm 0.03$ \\
\hline
Total &                      $ 32.1 \pm 0.1 $ & $ 21.3 \pm 0.08 $ & $ 1.45 \pm 0.05 $ \\
\end{tabular}
\end{ruledtabular}
\end{center}
\caption{Summary of tagging performance for the different tagging methods described in the text. Uncertainties shown are statistical only.  The efficiency and tagging power are each determined by summing over the individual bins of the charge distribution. The effective dilution is obtained from the measured efficiency and tagging power. The uncertainties are determined by combining the appropriate uncertainties on the individual bins of each charge distribution.}
\label{tab:taggingresults}
\end{table*}

The efficiency $\epsilon$ of an individual tagger is defined as the ratio of the number of tagged events to the total number of candidates. A probability that a specific event has a signal decay containing a ${\bar b}$-quark given the value of the discriminating variable  $P(B|Q)$ is constructed from the calibration samples for each of the $B^{+}$ and $B^{-}$ samples, defining $P(Q|B^{+})$ and $P(Q|B^{-})$ respectively. The probability to tag a signal event as containing a ${\bar b}$-quark is therefore $P(B|Q) = P(Q|B^{+}) / (P(Q|B^{+}) + P(Q|B^{-}))$ and  $P({\bar B}|Q) = 1-P(B|Q)$.   
The tagging power is defined as $\epsilon{\mathcal D}^2 = \sum_{i} \epsilon_{i} \cdot (2 P_{i}({B}|Q_{i}) - 1)^{2}$, where the sum is over the bins of the probability distribution as a function of the charge variable and $\epsilon_{i}$ is the number of tagged events in each bin divided by the total number of candidates.  An effective dilution ${\mathcal D}$ is calculated from the tagging power and the efficiency.

The combination of the tagging methods is applied according to the hierarchy of performance, based on the dilution of the tagging method. The single best performing tagging measurement is taken, according to the order: {\it combined} muon cone charge, {\it segment tagged} muon cone charge, jet charge. If it is not possible to provide a tagging response for the event, then a probability of 0.5 is assigned. A summary of the tagging performance is given in table 1.

\section{Maximum likelihood fit \label{sec:PDF}}
An unbinned maximum likelihood fit is performed on the selected events to extract the parameters of the \Bsto\ decay.  The fit uses information about the reconstructed mass $m$ and its  uncertainty $\sigma_m$, the measured proper decay time $t$ and its uncertainty  $\sigma_t$,  the tag probability, and the transversity angles $\Omega$ of each \Bst\ decay candidate.  There are three transversity angles; $\Omega = (\theta_T,\psi_T,\phi_T)$ and these are defined in section \ref{sec:sigPDF}.

The likelihood function is defined as a combination of the signal and background probability density functions as follows:
\begin{eqnarray} \label{eq:likelihood}
\ln~{\cal L}  & = & \sum_{i=1}^N \{  w_i \cdot \mathrm{ln} (  f_{\textrm{s}} \cdot   {\cal F}_{\textrm{s}}(  m_{i}, t_{i}, \Omega_{i},P(B|Q) ) \\ \nonumber
 & & +  f_{\textrm{s}} \cdot f_{B^0} \cdot {\cal F}_{B^0}  (  m_{i}, t_{i}, \Omega_{i},P(B|Q) )  \\ \nonumber
 & & + ( 1 -  f_{\textrm{s}} \cdot (1+ f_{B^0 } )    ) \cdot {\cal F}_{\textrm{bkg}} (  m_{i}, t_{i}, \Omega_{i},P(B|Q) ) \}  \\  \nonumber
\end{eqnarray}
where $N$ is the number of selected candidates, $w_i$ is a weighting factor to account for the trigger efficiency, $f_\textrm{s}$ is the fraction of signal candidates,  $f_{B^0}$ is the fraction of  $B^0$ ($\Bd \ra J/\psi K^{0*}$ and $\Bd \ra J/\psi K^{\pm}\pi^{\mp}$) mesons mis-identified as \Bs\ candidates calculated relative to the number of signal events; this parameter is  fixed  in the likelihood fit. The  mass $m_{i}$, the proper decay time $t_{i}$ and the decay angles $\Omega_{i} $ are the values measured from the data for each event $i$.  ${ \cal F}_{\textrm{s} }$, ${\cal F}_{B^0}$ and  ${\cal F}_{\textrm{bkg} }$  are the probability density functions (PDF) modelling the signal, the specific $B^0$  background and  the other background distributions, respectively.  A detailed description of the signal PDF terms in equation (\ref{eq:likelihood}) is given in section \ref{sec:sigPDF}.  The two background functions are, with the exception of new terms dependent on $P(B|Q)$ which are explained in section \ref{TagInFit}, unchanged from the previous analysis \cite{ATLASBsJpsiphi}.  They are each described by the product of eight terms which describe the distribution of each measured parameter.  With the exception of the lifetime and its uncertainty the background parameters are assumed uncorrelated.

\subsection{Signal PDF\label{sec:sigPDF}} 
The  PDF describing  the signal events, ${\cal F}_{\textrm{s}}$,   has the form of a product of PDFs for each quantity measured from the data:
\begin{eqnarray*} \label{eq:likelihoodSig}
  {\cal F}_{\textrm{s}}(  {\it m_{i}, t_{i},} \Omega_i, P(B|Q) ) & = & P_\textrm{s}(m_{i}, \sigma_{m_{i}}) \cdot P_\textrm{s}(\sigma_{m_{i}}) \nonumber \\
 & & \cdot P_\textrm{s}(\Omega_{i}, t_{i}, P(B|Q), \sigma_{t_{i}}) \cdot P_\textrm{s}(\sigma_{t_{i}}) \nonumber \\
 & & \cdot P_\textrm{s}(P(B|Q)) \nonumber \\
 & & \cdot A(\Omega_i,{\it p_{\mathrm{T}i}})  \cdot P_\textrm{s}({\it p_{\mathrm{T}i}})
 \end{eqnarray*}
 The terms $P_\textrm{s}(m_{i}, \sigma_{m_{i}})$, $P_\textrm{s}(\Omega_{i}, t_{i}, P(B|Q), \sigma_{t_{i}})$ and  $A(\Omega_i,{\it p_{\mathrm{T}i}})$ are explained in the current section. The tagging probability term $P_\textrm{s}(P(B|Q))$ is described in section \ref{TagInFit}.  The remaining probability terms $P_\textrm{s}(\sigma_{m_{i}})$, $P_\textrm{s}(\sigma_{t_{i}})$ and $P_\textrm{s}({\it p_{\mathrm{T}i}})$ are described by Gamma functions.  They are unchanged from the previous analysis and explained in detail in ref. \cite{ATLASBsJpsiphi}. Ignoring detector effects, the joint distribution for the decay time $t$ and the transversity angles $\Omega$ for the \Bsto\ decay is given by the differential decay rate \cite{CDFAngles}:
\begin{equation*}\label{eq:ang2}
\frac{d^4 \Gamma}{dt\ d\Omega}= \sum_{k=1}^{10} {\cal O}^{(k)}(t) g^{(k)}(\theta_T,\psi_T,\phi_T), 
\end{equation*}
where  ${\cal O}^{(k)}(t)$ are the time-dependent amplitudes and $g^{(k)}(\theta_T,\psi_T,\phi_T)$ are the angular functions, given in table \ref{tab:SigPDF}.  The formulae for the time-dependent amplitudes have the same structure for \Bs\ and \aBs\ but with a sign reversal in the terms containing $\Delta m_s$.  The addition of flavour tagging to the analysis means that these terms no longer cancel, so there are more terms in the fit that contain \phis.  In addition to this, the strong phase variable $\delta_\perp$ becomes accessible and one of the symmetries in the untagged fit is removed.  $A_\perp(t)$ describes a $CP$ odd final-state configuration while both $A_0(t)$ and $A_\parallel(t)$ correspond to $CP$ even final-state configurations.   $A_S(t)$ describes the contribution of the $CP$ odd non-resonant $\Bs \rightarrow J/\psi K^+K^-$ $S$-wave state  as well as the  $\Bs \rightarrow J/\psi f_{0}$   decays.  The corresponding amplitudes are given in the last four lines of  table \ref{tab:SigPDF}  ($k=7$--$10$) and follow the convention used in the previous analysis~\cite{LHCb:2011aa}. The likelihood is independent of the $K^+K^-$ mass distribution. 

\begin{table*}
\begin{center}
\begin{ruledtabular}
\begin{tabular}{ c| l | l }
$k$ & ${\cal O}^{(k)}(t)$ & $g^{(k)}(\theta_T,\psi_T,\phi_T)$ \\
\hline \hline
1 & $\frac{1}{2} |A_0(0)|^2 \left[\left(1+\cos\phis\right) e^{-\Gamma_{\rm L}^{(s)} t} + \left(1-\cos\phis\right) e^{-\Gamma_{\rm H}^{(s)} t} \pm 2 e^{-\Gamma_st} \sin (\Delta m_s t)  \sin \phi_s \right]$ & $2 \cos^2\psi_T(1 - \sin^2\theta_T\cos^2\phi_T)$ \\
2 & $\frac{1}{2}|A_{\|}(0)|^2 \left[\left(1+\cos\phis\right) e^{-\Gamma_{\rm L}^{(s)} t} + \left(1-\cos\phis\right) e^{-\Gamma_{\rm H}^{(s)} t} \pm 2 e^{-\Gamma_st} \sin (\Delta m_s t)  \sin \phi_s \right]$ & $\sin^2\psi_T(1 - \sin^2\theta_T\sin^2\phi_T)$ \\
3 & $\frac{1}{2} |A_{\perp}(0)|^2 \left[\left(1- \cos\phis\right)e^{-\Gamma_{\rm L}^{(s)} t} + \left(1+\cos\phis\right)e^{-\Gamma_{\rm H}^{(s)} t} \mp 2 e^{-\Gamma_st} \sin (\Delta m_s t)  \sin \phi_s \right]$ & $\sin^2\psi_T\sin^2\theta_T$ \\
4 & $\frac{1}{2}|A_0(0)||A_{\|}(0)|\cos\delta_{||} $ & $- \frac{1}{\sqrt{2}}\sin2\psi_T\sin^2\theta_T\sin2\phi_T $ \\
  & \hfill $\left[\left(1+\cos\phis\right)e^{-\Gamma_{\rm L}^{(s)} t} + \left(1-\cos\phis\right)e^{-\Gamma_{\rm H}^{(s)} t} \pm 2 e^{-\Gamma_st} \sin (\Delta m_s t)  \sin \phi_s\right]$&  \\
5 & $|A_{\|}(0)||A_{\perp}(0)| [  \frac{1}{2}(e^{-\Gamma_{\rm L}^{(s)} t}-e^{-\Gamma_{\rm H}^{(s)} t})\cos(\delta_{\perp} - \delta_{||})\sin\phis$ & $\sin^2\psi_T\sin2\theta_T\sin\phi_T$ \\
 & \hfill $\pm e^{-\Gamma_st}(\sin (\delta_\perp -\delta_\|) \cos(\Delta m_s t) - \cos (\delta_\perp - \delta_\|)\cos \phi_s \sin (\Delta m_st) ) ]$ & \\
 6 & $|A_{0}(0)||A_{\perp}(0)| [  \frac{1}{2}(e^{-\Gamma_{\rm L}^{(s)} t}-e^{-\Gamma_{\rm H}^{(s)} t})\cos \delta_{\perp} \sin\phis$ & $\frac{1}{\sqrt{2}} \sin2\psi_T\sin2\theta_T\cos\phi_T$ \\
 & \hfill $\pm e^{-\Gamma_st}(\sin \delta_\perp \cos(\Delta m_s t) - \cos \delta_\perp \cos \phi_s \sin (\Delta m_st) ) ]$ & \\
7 &$\frac{1}{2} |A_{S}(0)|^2 \left[\left(1- \cos\phis\right)e^{-\Gamma_{\rm L}^{(s)} t} + \left(1+\cos\phis\right)e^{-\Gamma_{\rm H}^{(s)} t} \mp 2 e^{-\Gamma_st} \sin (\Delta m_s t)  \sin \phi_s \right]$ &$ \frac{2}{3}\left(1-\sin^2\theta_T\cos^2\phi_T\right) $ \\
 8 & $|A_{S}(0)||A_{\parallel}(0)| [  \frac{1}{2}(e^{-\Gamma_{\rm L}^{(s)} t}-e^{-\Gamma_{\rm H}^{(s)} t})\sin (\delta_\| - \delta_S) \sin\phis$ & $\frac{1}{3} \sqrt{6} \sin\psi_T\sin^2\theta_T\sin 2\phi_T$ \\
 & \hfill $\pm e^{-\Gamma_st}(\cos ( \delta_\| - \delta_S ) \cos(\Delta m_s t) - \sin ( \delta_\| - \delta_S ) \cos \phi_s \sin (\Delta m_st) ) ]$ & \\
9 &$\frac{1}{2}|A_{S}(0)| |A_{\perp}(0)|\sin(\delta_{\perp}-\delta_{S}) $ & $\frac{1}{3} \sqrt{6} \sin\psi_T\sin 2\theta_T\cos\phi_T$ \\
  &\hfill $\left[\left(1- \cos\phis\right)e^{-\Gamma_{\rm L}^{(s)} t} + \left(1+\cos\phis\right)e^{-\Gamma_{\rm H}^{(s)} t} \mp 2 e^{-\Gamma_st} \sin (\Delta m_s t)  \sin \phi_s \right]$ &  \\
 10 & $|A_{0}(0)||A_{S}(0)| [  \frac{1}{2}(e^{-\Gamma_{\rm H}^{(s)} t}-e^{-\Gamma_{\rm L}^{(s)} t})\sin  \delta_S \sin\phis$ & $\frac{4}{3}\sqrt{3}\cos\psi_T\left(1-\sin^2\theta_T\cos^2\phi_T\right)$ \\
 & \hfill $\pm e^{-\Gamma_st}(\cos \delta_S \cos(\Delta m_s t) + \sin \delta_S \cos \phi_s \sin (\Delta m_st) ) ]$ & \\
\end{tabular}
\end{ruledtabular}
\end{center}
\caption{Table showing the ten time-dependent amplitudes, ${\cal O}^{(k)}(t)$ and the functions of the transversity angles $g^{(k)}(\theta_T,\psi_T,\phi_T)$. The amplitudes $|A_0(0)|^2$ and $|A_\parallel(0)|^2$ are for the $CP$ even components of the \Bst\ decay,  $|A_{\perp}(0)|^2$ is the $CP$ odd amplitude; they have corresponding strong phases $\delta_{0}$, $\delta_{\parallel}$ and $\delta_{\perp}$, by convention $\delta_0$ is set to be zero.  The $S$-wave amplitude $|A_S(0)|^2$ gives the fraction of \BstoKK\ and has a related strong phase $\delta_S$. The $\pm$ and $\mp$ terms denote two cases: the upper sign describes the decay of a meson that was initially a \Bs, while the lower sign describes the decays of a meson that was initially \aBs. \label{tab:SigPDF}}
\end{table*}

The equations are normalized, such that the squares of the amplitudes sum to unity; three of the four amplitudes are fit parameters and  $|A_\perp(0)|^2$ is determined according to this constraint.

The angles ($\theta_T,\psi_T,\phi_T$), are defined in the rest frames of the final-state particles.  The $x$-axis is determined by the direction of the $\phi$ meson in the $J/\psi$ rest frame, and the $K^{+}K^{-}$  system defines the $x$--$y$ plane, where $p_{y}(K^{+})>0$. The three angles are defined as follows:
\begin{itemize}
\item$\theta_T$, the angle between $\vec{p}(\mu^{+})$ and the normal to the $x$--$y$ plane, in the $J/\psi$ meson rest frame
\item $\phi_T$, the angle between the $x$-axis and $\vec{p}_{xy}(\mu^{+})$, the projection of the $\mu^+$ momentum in the $x$--$y$ plane, in the $J/\psi$ meson rest frame
\item $\psi_T$, the angle between $\vec{p}(K^{+})$ and $-\vec{p}(J/\psi)$ in the $\phi$ meson rest frame 
\end{itemize}

The signal PDF, $P_s(\Omega, t,P(B|Q), \sigma_{t})$, needs to take into account lifetime resolution, so each time element in table \ref{tab:SigPDF} is smeared with a Gaussian function. This smearing is done numerically on an event-by-event basis where the width of the Gaussian function is the proper decay time uncertainty, measured for each event, multiplied by a scale factor to account for any mis-measurements. 

The angular sculpting of the detector and kinematic cuts on the angular distributions are included in the likelihood function through $A(\Omega_i, {\it p_{Ti}})$. This is calculated using a 4D binned acceptance method, applying an event-by-event efficiency according to the transversity angles ($\theta_T,\psi_T,\phi_T$) and the \pT\ of the candidate. The \pT\  binning is necessary, because the angular sculpting is influenced by the \pT\  of the \Bs.  The acceptance was calculated from the \Bst\ MC events. In the likelihood function, the acceptance is treated as an angular sculpting PDF, which is multiplied with the time- and angular-dependent PDF describing the \Bsto\ decays.  As both the acceptance and time-angular decay PDFs depend on the transversity angles they must be normalized together. This normalization is done numerically during the likelihood fit. 

The signal mass function, $P_s(m)$, is modelled using a single Gaussian function smeared with an event-by-event mass resolution. The PDF is normalized over the range $5.15 < m(\Bs) < 5.65 \GeV$.

\subsection{Using tag information in the fit} \label{TagInFit}
The tag probability for each \Bs\ candidate is determined from a weighted sum of charged-particle tracks in a cone, as described in section~\ref{sec:Tagging}.  The tag probability is obtained from this tag charge using the calibrations measured in the $B^{\pm}\to J/\psi K^{\pm}$ data.  For the case where there is only one track, the cone charge can only be $\pm 1$.  This leads to a tag probability distribution with continuous and discrete parts (spikes), which are estimated separately.  The distributions of tag probabilities for the signal and background are also different and  since the background  cannot be factorized out, extra PDF terms are included to account for this difference.

\begin{figure}
	\begin{center}
		\includegraphics[width=0.43\textwidth]{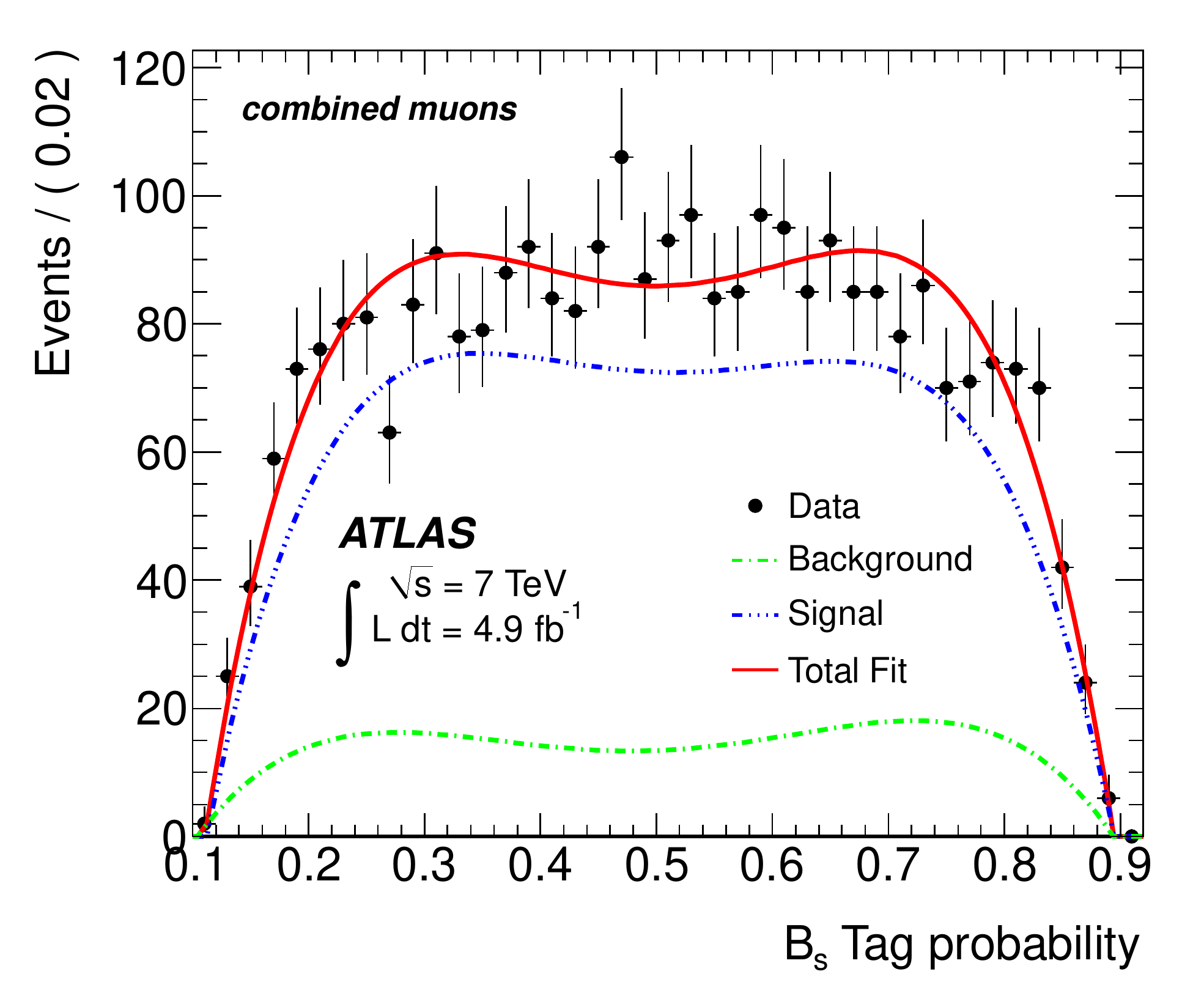}
		\includegraphics[width=0.43\textwidth]{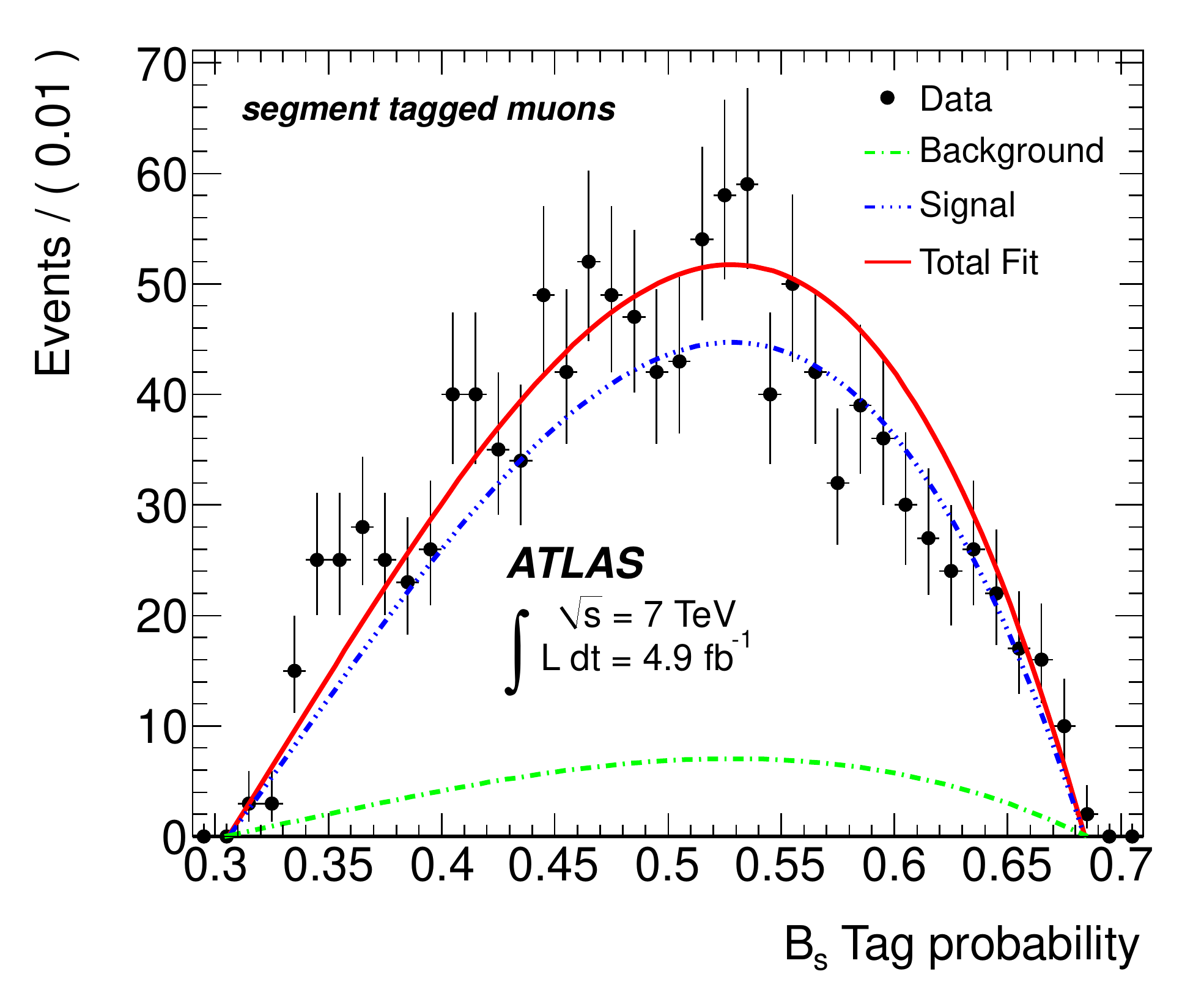}
		\includegraphics[width=0.43\textwidth]{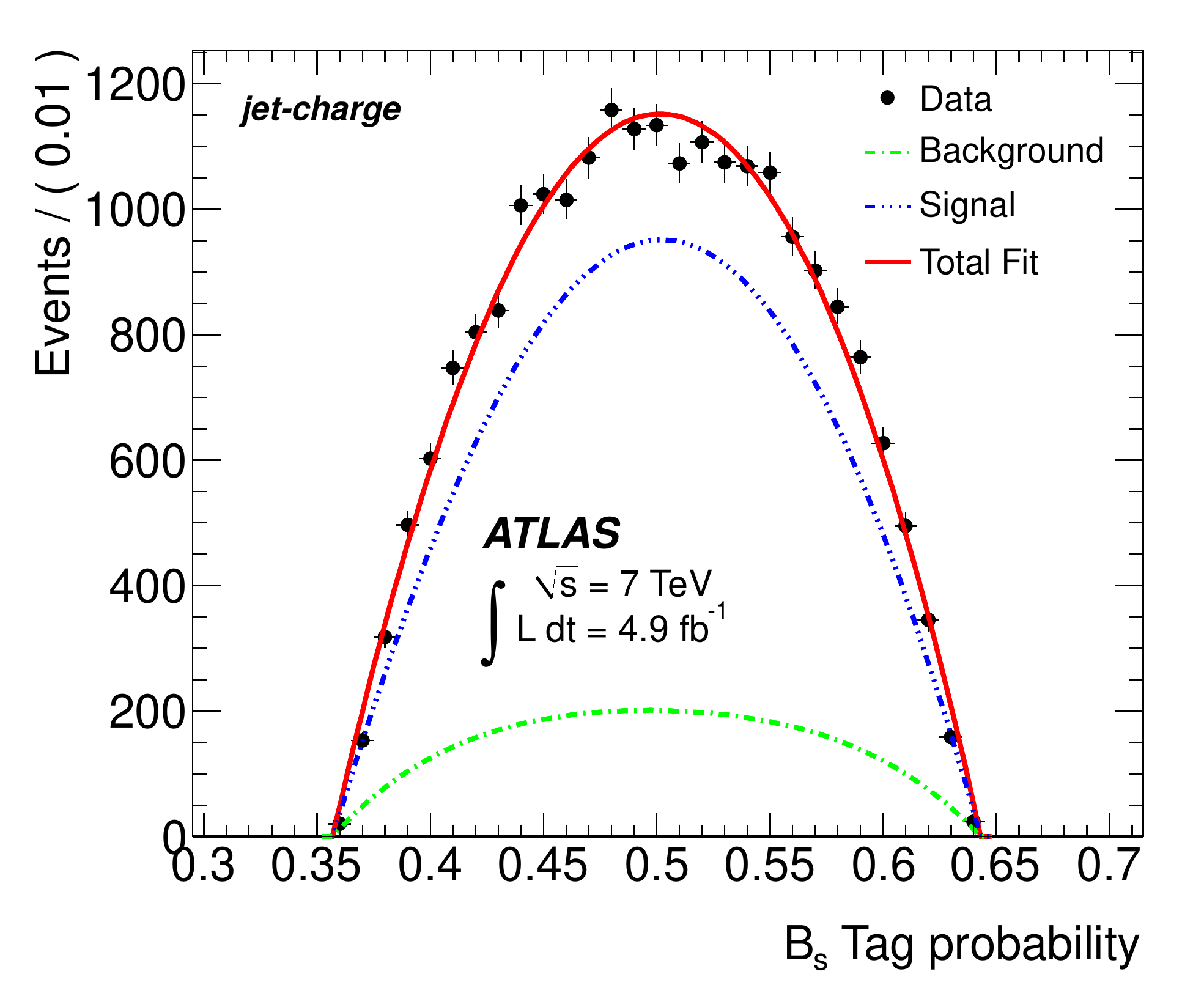}
		\caption{The \Bs-tag probability distribution for the events tagged with combined muons (top), segment tagged muons (middle) and jet charge (bottom). Black dots are data after removing spikes, blue is the fit to the sidebands, green to the signal and red is a sum of both fits.\label{Fig:taggerfit}}
	\end{center}	
\end{figure}

To describe the continuous parts, the sidebands are parameterized first. Sidebands are selected  according to \Bs\ mass, i.e. $m(\Bs) <5.317$~GeV or $m(\Bs)>5.417$~GeV.   In the  fit the same function as for the sidebands is  used to describe events in the  signal region: background  parameters are fixed to the values obtained in sidebands while signal parameters are free in this step.   The ratio of  background to signal (obtained from simultaneous mass--lifetime fit) is fixed as well.   The function describing tagging using combined muons has the form of a fourth-order Chebychev polynomial. A third-order polynomial is used for the segment tagged muons' tagging algorithm.  A fourth-order Chebychev polynomial is also applied for the jet charge tagging algorithm.   In all three cases unbinned maximum likelihood fits are used. Results of fits projected on histograms are shown in figure~\ref{Fig:taggerfit}. 

The spikes have their origin in tagging objects formed from a single track, providing a tag charge of exactly $+1$ or $-1$. When a background candidate is formed from a random combination of a $J/\psi$  and a pair of tracks, the positive and negative charges are equally probable. However, some of the background events are formed of  partially reconstructed $B$ hadrons, and in these cases tag charges of $+1$ or $-1$ are not equally probable. For signal events the tag charges are obviously not symmetric. The fractions $f_{+1}$ and  $f_{-1}$ of events tagged with charges of $+1$ and $-1$ are derived separately for signal and background.  The remaining $(1- f_{+1} -f_{-1})$  is the fraction of events in the continuous region. The  fractions $f_{+1}$  and $f_{-1}$ are determined using the same \Bs\ mass sidebands and signal regions as in case of continuous parts. Table \ref{tab:TagPunziSpkes} summarizes the obtained relative probabilities between tag charges $+1$ and $-1$ for signal and background events and for all tag methods.

\begin{table*}
\begin{center}
\begin{ruledtabular}
\begin{tabular}{l|cc|cc}
  Tag method & \multicolumn{2}{c|}{Signal} & \multicolumn{2}{c}{Background} \\
             & $f_{+1}$ & $f_{-1}$         & $f_{+1}$ & $f_{-1}$ \\
  \hline \hline
Combined $\mu$ & $0.106 \pm 0.019$ & $0.187 \pm 0.022$ & $0.098 \pm 0.006$ & $0.108 \pm 0.006$ \\
Segment tag $\mu$      & $0.152 \pm 0.043$ & $0.153 \pm 0.043$ & $0.098 \pm 0.009$ & $0.095 \pm 0.008$ \\
Jet charge     & $0.167 \pm 0.010$ & $0.164 \pm 0.010$ & $0.176 \pm 0.003$ & $0.180 \pm 0.003$ \\
\end{tabular}
\end{ruledtabular}
\end{center}
\caption{Table summarizing the obtained relative probabilities between tag charges $+1$ and $-1$ for signal and background events for the different tagging methods. Only statistical errors are quoted.  The asymmetry in the signal combined-muon tagging method has no impact on the results as it  affects only $1\%$ of the signal events (in addition to the negligible effect of the tag-probability distributions themselves).}
\label{tab:TagPunziSpkes}
\end{table*}

\begin{table*}
\begin{center}
\begin{ruledtabular}
\begin{tabular}{l|c|c}
  Tag method & Signal      & Background \\
  \hline\hline
  Combined $\mu$ & $0.0372 \pm 0.0023$ & $0.0272 \pm 0.0005$ \\
Segment tag $\mu$      & $0.0111 \pm 0.0014$ & $0.0121 \pm 0.0003$ \\
Jet charge     & $0.277  \pm 0.007$  & $0.254  \pm 0.002$ \\
Untagged       & $0.675  \pm 0.011$  & $0.707  \pm 0.003$ \\
\end{tabular}
\end{ruledtabular}
\end{center}
\caption{Table summarizing the relative population of the tagging methods in the background and signal events. Only statistical errors are quoted.}
\label{tab:TagPunziRelat}
\end{table*}

Similarly, the sideband subtraction method is also used to determine, for signal and background events, the relative fraction of each tagging method. The results are summarized in table \ref{tab:TagPunziRelat}.

If the tag-probability PDFs were ignored in the likelihood fit, equivalent to assuming identical signal and background behaviour, the impact on the fit result would be small, affecting the results by less than $10\%$ of the statistical uncertainty.

\section{Results\label{sec:results}}
The full simultaneous maximum likelihood fit contains 25 free parameters.  These include the nine physics parameters: \DGs, \phis, \Gs, $|A_{0}(0)|^2$, $|A_{\parallel}(0)|^2$, $\delta_{||}$, $\delta_{\perp}$, $|A_{S}|^2$ and $\delta_{S}$.  The other parameters in the likelihood function are the \Bs\ signal fraction $f_s$, the parameters describing the $J/\psi \phi$ mass distribution, the parameters describing the \Bs\ meson decay time plus angular distributions of background events, the parameters used to describe the estimated decay time uncertainty distributions for signal and background events, and scale factors between the estimated decay time and mass uncertainties and their true uncertainties. 

\begin{figure}
\begin{center}
\includegraphics[width=0.46\textwidth]{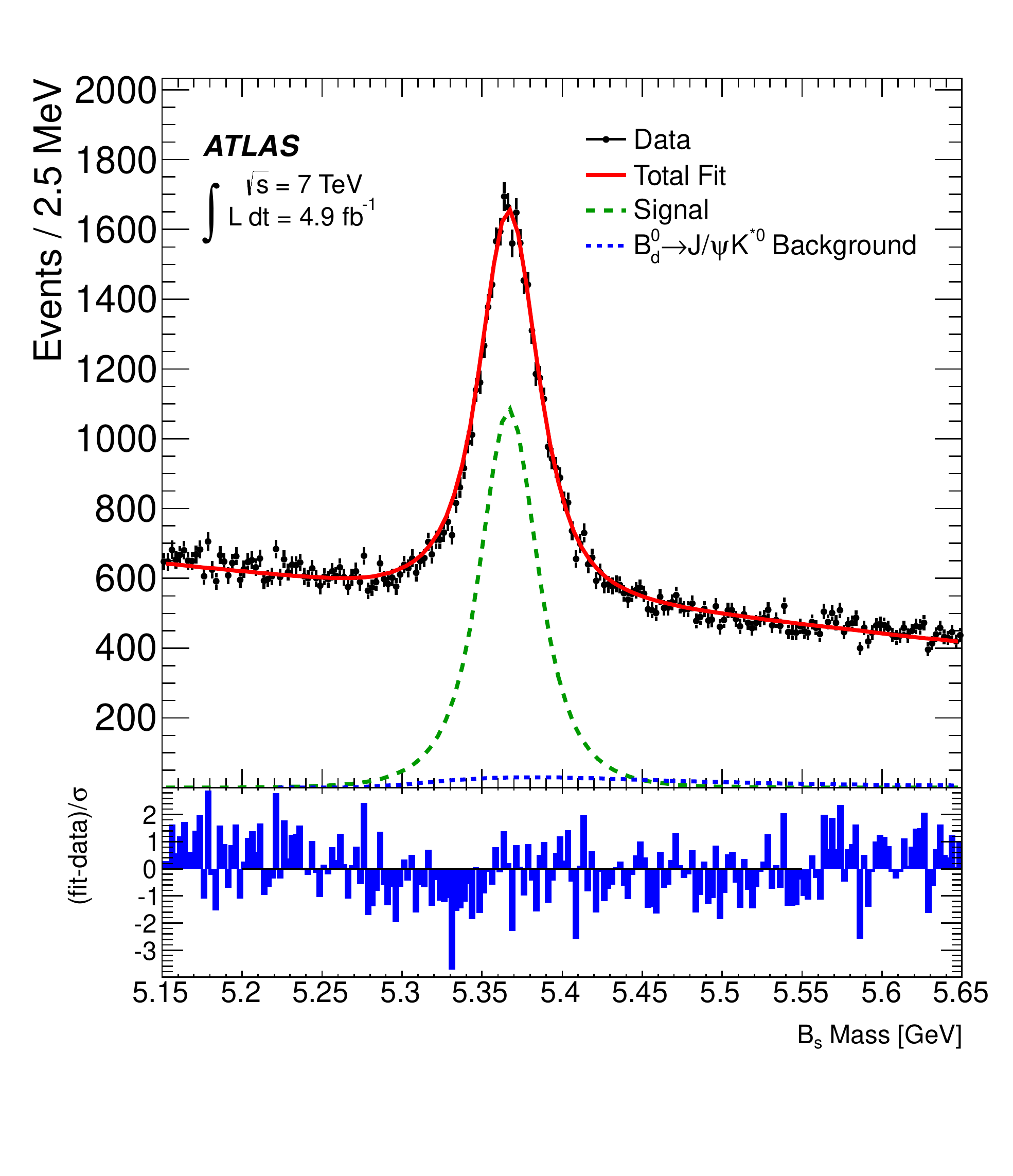}
\includegraphics[width=0.46\textwidth]{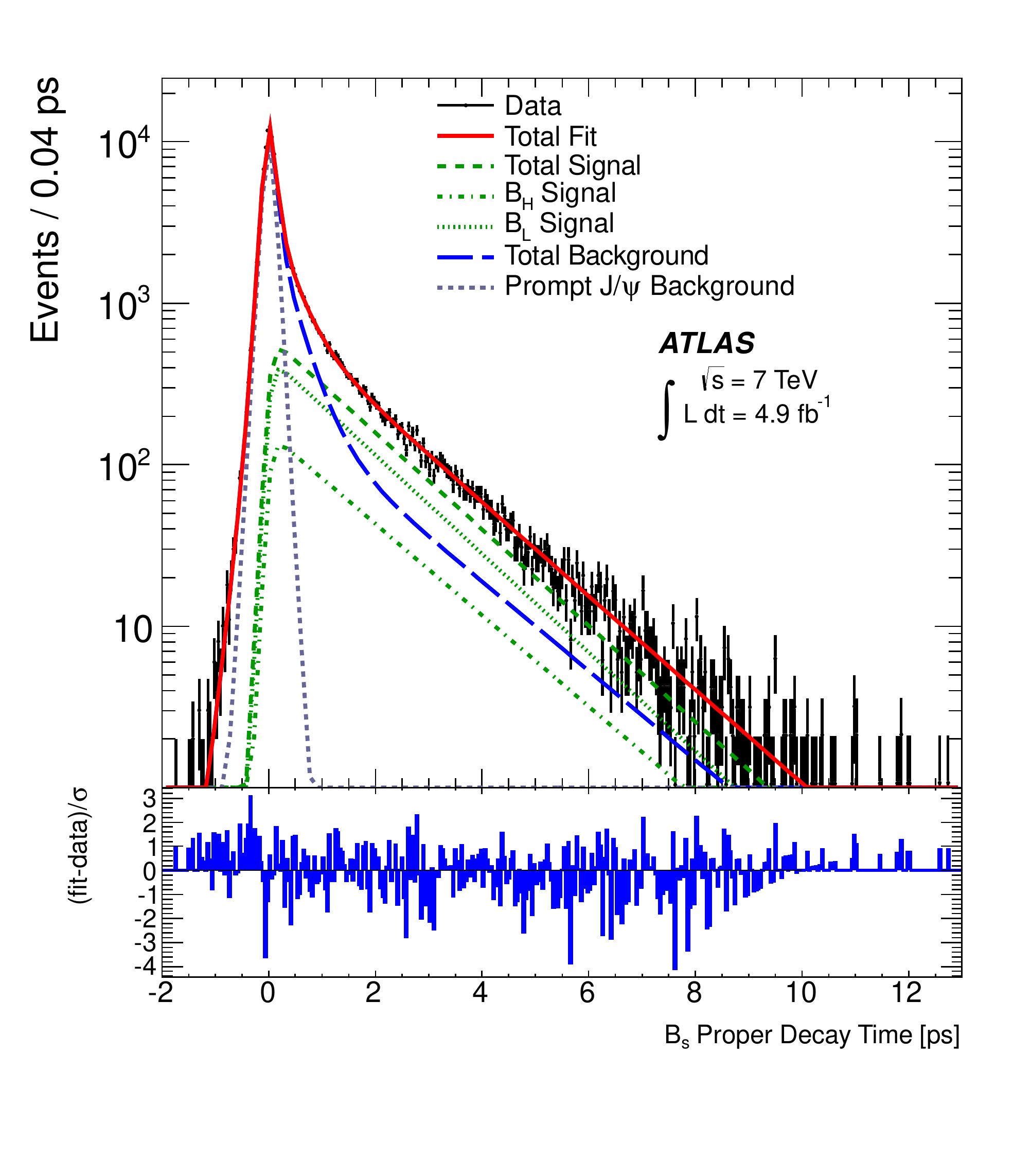}
\caption{(Top) Mass fit projection for the \Bst. The red line shows the total fit, the dashed green line shows the signal component while the dotted blue line shows the contribution from $\Bd \ra J/\psi K^{0*}$ events. (Bottom) Proper decay time fit projection for the \Bst. The red line shows the total fit while the green dashed line shows the total signal.  The light and heavy components of the signal are shown in green as a dotted and a dash-dotted line, respectively.  The total background is shown as a blue dashed line with a grey dotted line showing the prompt $J/\psi$ background.  The pull distributions at the bottom show the difference between data and fit value normalized to the data statistical uncertainty.}
\label{fig:simulFit}
\end{center}
\end{figure}

\begin{figure}
\begin{center}
\includegraphics[width=0.43\textwidth]{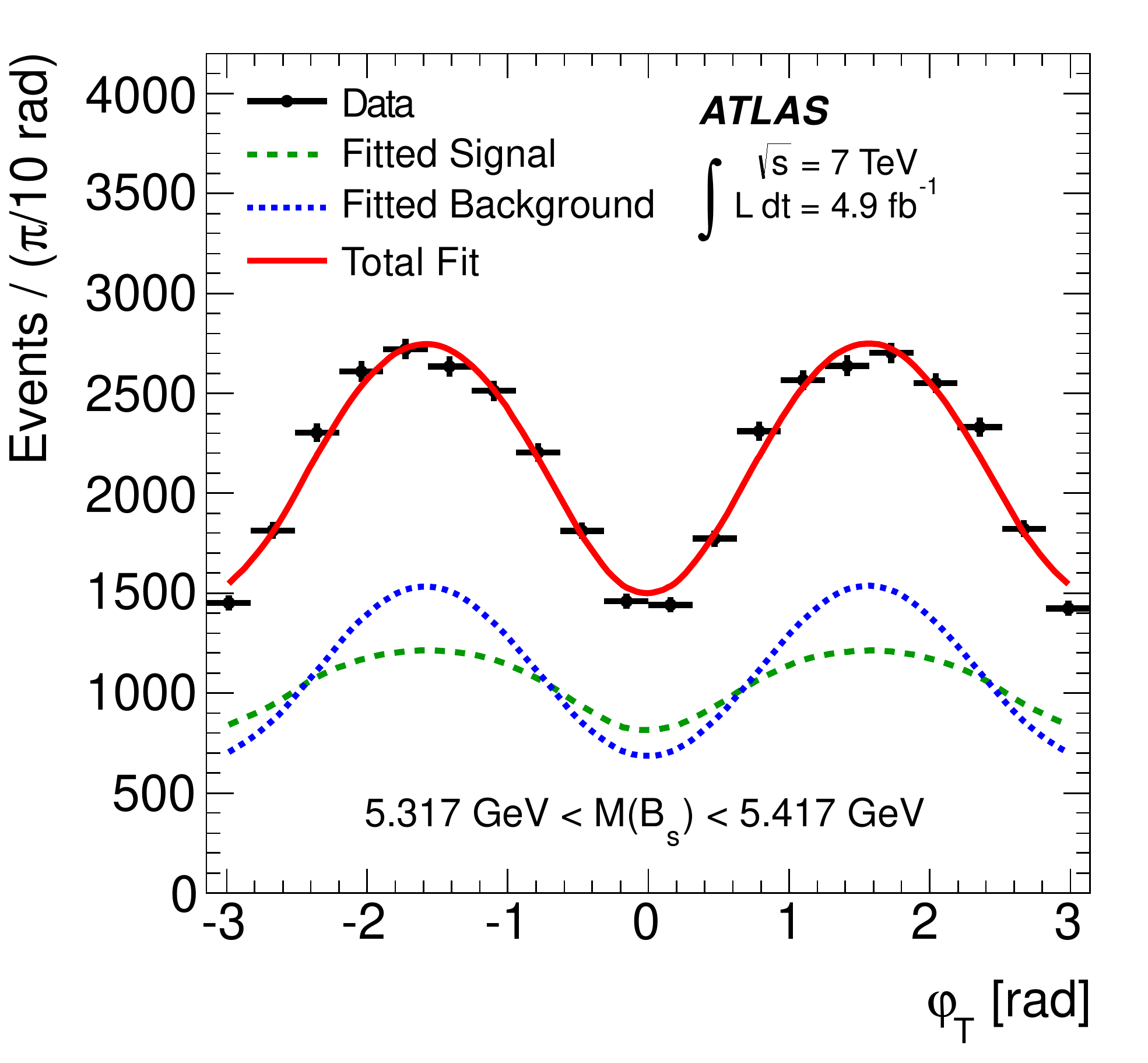}
\includegraphics[width=0.43\textwidth]{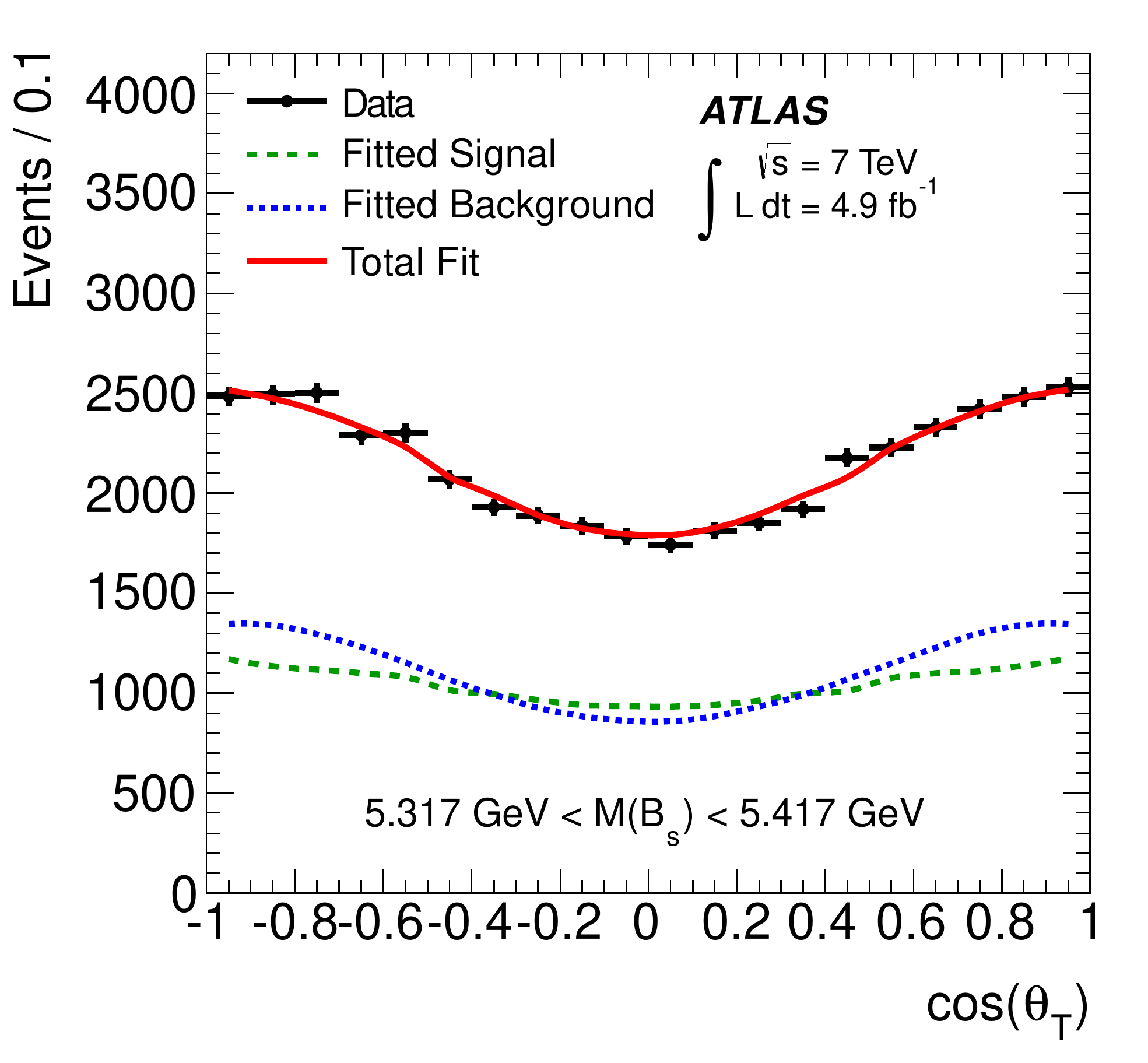}
\includegraphics[width=0.43\textwidth]{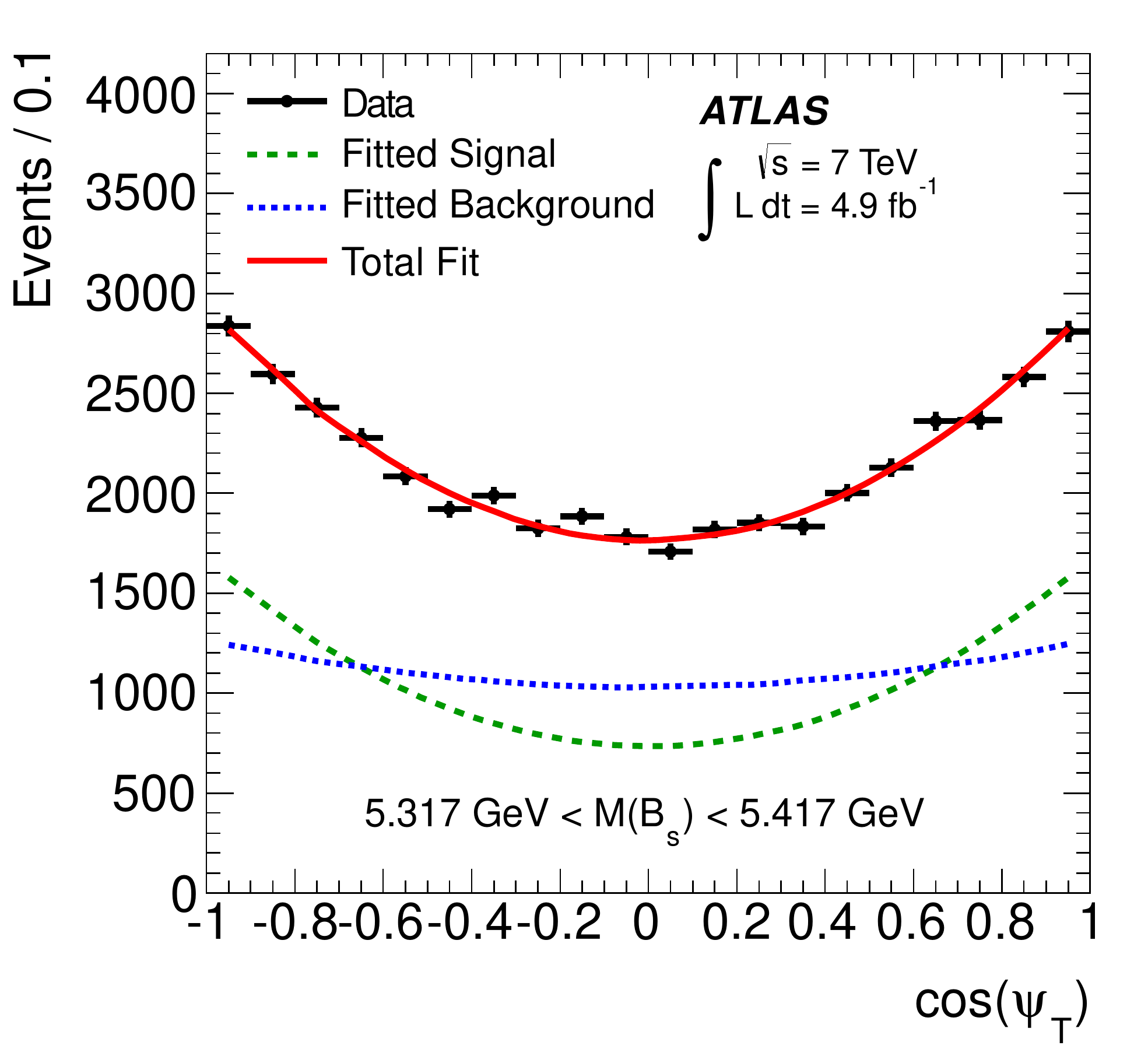}
\caption{Fit projections for transversity angles. (Top): $\phi_T$, (middle): $\cos\theta_T$, (bottom): $\cos\psi_T$.  In all three plots, the red line shows the total fit, the dashed green line shows the signal component and the dotted blue line shows the background contribution.}
\label{fig:AngularProjections}
\end{center}
\end{figure}

The number of signal \Bs\ meson candidates extracted from the fits is \nsig. The results and correlations for the measured physics parameters of the simultaneous unbinned maximum likelihood fit are given in tables~\ref{tab:FitResults} and~\ref{tab:FitResultsCor}.  Fit projections of the mass, proper decay time and angles are given in figures~\ref{fig:simulFit} and \ref{fig:AngularProjections} respectively.

\begin{table}
\begin{center}
\begin{ruledtabular}
\begin{tabular}{c|c|c|c}
Parameter & Value & Statistical & Systematic  \\
          &       & uncertainty & uncertainty \\
\hline\hline
 \phis [rad] & \pSfit & \pSstat & \pSsyst \\
 \DGs [ps$^{-1}$] & \DGfit & \DGstat & \DGsyst \\
 \Gs [ps$^{-1}$] & \GSfit & \GSstat & \GSsyst \\
  $|A_{\parallel}(0)|^2$ & \Apasqfit & \Apasqstat & \Apasqsyst \\
  $|A_{0}(0)|^2$ & \Azesqfit  & \Azesqstat & \Azesqsyst \\
$|A_{S}(0)|^2$ & \ASsqfit & \ASsqstat & \ASsqsyst \\
$\delta_\perp$ & \Deltperpfit  & \Deltperpstat & \Deltperpsyst \\
$\delta_{\parallel}$  &\multicolumn{2}{c|}{\Deltparafit} &\Deltparasyst \\
$\delta_{\perp} -\delta_{S}$ & \multicolumn{2}{c|}{\DeltPSfit} & \DeltPSsyst  \\
\end{tabular}
\end{ruledtabular}
\end{center}
\caption{Fitted values for the physical parameters with their statistical and systematic uncertainties. For the parameters $\delta_{\parallel}$ and $\delta_{\perp} -\delta_{S}$ a $68\%$ confidence level interval is given.  The reason for this is described in section 8.\label{tab:FitResults}}
\end{table}

\clearpage

\begin{table*}
\begin{center}
\begin{ruledtabular}
\begin{tabular}{c|c|c|c|c|c|c|c|c|c}
		& $\phi_{s}$ 	& $\Delta\Gamma$	& $\Gamma_{s}$ 	& $|A_{||}(0)|^2$ 	& $|A_0(0)|^2$ 	& $|A_S(0)|^2$ 	& $\delta_{\parallel}$  & $\delta_{\perp}$  & $\delta_{\perp} -\delta_{S} $ \\	
\hline \hline
$\phi_{s}$ 		& 1.000 	& \pDG		& \pG 	& \pApa 	 	& \pAz 	& \pAs 	& \pDpara     & \pDperp    &  \pDpS  \\
$\Delta\Gamma$ 		&      & 1.000 		& \DGG	&  \DGApa 	& \DGAz 	& \DGAs 	& \DGDpara  & \DGDperp & \DGDpS \\
$\Gamma_{s}$ 		& 		&  			& 1.000 	& \GApa     	& \GAz 	& \GAs 	& \GDpara     & \GDperp   & \GDpS \\
 $|A_{||}(0)|^2$ 	        &  		&		  	&  		& 1.000 		&  \AzApa & \ApaAs 	& \ApaDpara  & \ApaDperp  & \ApaDpS \\
 $|A_0(0)|^2$ 		&  		&  			&  		&  			& 1.000 	& \AzAs 	& \AzDpara  & \AzDperp & \AzDpS \\
  $|A_S(0)|^2$ 		&  		&  			&  		&  			&  		& 1.000 	& \AsDpara  & \AsDperp & \AsDpS\\
   $\delta_{\parallel}$ &  		&  			&  		&  			&  		& 	 	& 1.000  & \DparaDperp & \DparaDpS \\
 $\delta_{\perp}$   &  		&  			&  		&  			&  		& 	 	& & 1.000   &  \DperpDpS \\
 $\delta_{\perp} -\delta_{S} $&  		&  			&  		&  			&  		& 	 	& & & 1.000 \\
\end{tabular}
\end{ruledtabular}
\end{center}
\caption{ Correlations between the physics parameters. \label{tab:FitResultsCor}}
\end{table*}

\section{Systematic uncertainties}
Systematic uncertainties are assigned by considering several effects that are not accounted for in the likelihood fit. These are described below.  
\begin{itemize}
\item {\bf Inner detector alignment:} Residual mis-alignments of the inner detector affects the impact parameter distribution with respect to the primary vertex. The effect of the residual mis-alignment is estimated using simulated events with and without distorted geometry. For this, the impact parameter distribution with respect to the primary vertex is measured with data as a function of $\eta$ and $\phi$ with the maximum deviation from zero being less than 10 $\mu$m. The measurement is used to distort the geometry for simulated events  in order to reproduce the impact parameter distribution measured as a function of $\eta$ and $\phi$. The difference between the measurement using simulated events with and without the distorted geometry is used as the systematic uncertainty. 

\item {\bf Trigger efficiency:}  It is observed that the muon trigger biases the transverse impact parameter of muons toward smaller values.  To correct for this bias the events are re-weighted according to 
\begin{equation}
   w =  e^{   -|t|/ ( \tau_{\mathrm{sing}}+\epsilon )     }   /     e^{   -|t|/ \tau_{\mathrm{sing}}    } ,    \nonumber
\end{equation}
where $ \tau_{\mathrm{sing}}$ is a single \Bs\ lifetime measured before the correction, using an unbinned mass--lifetime maximum likelihood fit. The value of the  parameter $\epsilon$ and its uncertainty are described in   \cite{ATLASBsJpsiphi}. The systematic uncertainty is calculated by varying the value of $\epsilon$ by its uncertainty and rerunning the fit.

\item  {\bf \boldmath{$B^0$} contribution:} Contaminations from $B^0 \ra J/\psi K^{0*}$ and $B^0 \ra J/\psi K\pi$ events mis-reconstructed as \Bst\ are accounted for in the default fit.  The fractions of $B^0 \ra J/\psi K^{0*}$ and $B^0 \ra J/\psi K\pi$ events in the default fit are $(6.5 +/- 2.4)\%$ and $(4.5 +/- 2.8)\%$ respectively. They were determined in MC simulation and using branching fractions from ref. \cite{PDG}. To estimate the systematic uncertainty arising from the precision of the fraction estimates, the data are fitted with these fractions increased and decreased by 1$\sigma$.  The largest shifts in the fitted values from the default case are taken as the systematic uncertainty for each parameter of interest. 

\item {\bf Tagging:} For the uncertainties in the fit parameters due to uncertainty in the tagging, the statistical and systematic components are separated. The statistical uncertainty is due to the sample size of $B^\pm \ra J/\psi K^\pm$ decays available and is included in the overall statistical error. The systematic uncertainty arises from the precision of the tagging calibration and is estimated by varying the model parameterizing the probability distribution, $P(B|Q)$, as a function of tag charge. The default model is a linear function. For the combined-muon cone-charge tag and the segment tagged muons the alternative fit function is a third-order polynomial.  For the jet-charge tag with no muons, a third- and a fifth-order polynomial are used.  The \Bs\ fit was repeated using the alternative models and the largest difference was assigned as the systematic uncertainty.

\item {\bf Angular acceptance method:} The angular acceptance is calculated from a binned fit to Monte Carlo data.   A separate set of Monte Carlo signal events were generated and fully simulated.   Background was generated using pseudo-experiments as described below.  There is sufficient data to perform 166 fits.  The systematic uncertainty is calculated using the bias of the pull distribution multiplied by the statistical uncertainty of each parameter. In order to estimate the size of the systematic uncertainty introduced from the choice of binning, different acceptance functions are calculated using different bin widths and central values. These effects are found to be negligible. 

\item {\bf Signal and background mass model, resolution model, background lifetime and background angles model:} In order to estimate the size of systematic uncertainties caused by the assumptions made in the fit model, variations of the model are tested in pseudo-experiments. A set of 2400 pseudo-experiments is generated for each variation considered, and fitted with the default model.   The systematic error quoted for each effect is the difference between the mean shift of the fitted value of each parameter from its input value for the pseudo-experiments with the systematic alteration included. The variations are: two different scale factors are used to generate the signal mass; the background mass is generated from an exponential function; two different scale factors are used to generate the lifetime uncertainty; the background lifetimes are generated by sampling data from the mass sidebands; pseudo-experiments are generated with background angles taken from histograms from sideband data and are fitted with the default fit model to assess the systematic uncertainty to the parameterization of the background angles in the fit.  

\item {\bf Default fit model:} The systematic uncertainty of the default fit model is calculated using the bias of the pull distribution of 2400 pseudo-experiments, multiplied by the statistical uncertainty of each parameter.

\end{itemize}
The systematic uncertainties are provided in table~\ref{tab:syst_totals}. For each variable, the total systematic error is obtained by adding in quadrature the different contributions.

\begin{table*}
\begin{center}
\begin{ruledtabular}
\begin{tabular}{l| ccc ccc ccc} 
& \phis  & \DGs & \Gs   &  $|A_{\parallel}(0)|^2$ & $|A_{0}(0)|^2$ & $ |A_{S}(0)|^2 $ &  $\delta_{\perp}$& $\delta_{\parallel}$  & $\delta_{\perp} -\delta_{S} $	\\  
&[rad]&[ps$^{-1}$]&[ps$^{-1}$]& &&& [rad]&[rad]&  [rad] \\
\hline
ID alignment    &${\small<}10^{-2}$            &${\small<}10^{-3}$      &${\small<}10^{-3}$          &${\small<}10^{-3}$        &${\small<}10^{-3}$      &  - &${\small<}10^{-2}$   &${\small<}10^{-2}$   &  - \\
Trigger efficiency                &${\small<}10^{-2}$     &${\small<}10^{-3} $    &   0.002        &   ${\small<}10^{-3} $    &   ${\small<}10^{-3} $  & $<10^{-3}$     &${\small<}10^{-2}$   &${\small<}10^{-2}$   & ${\small<}10^{-2}$\\
$\Bd$ contribution               & 0.03            & $  0.001$       &${\small<}10^{-3}$   &${\small<}10^{-3}$       & 0.005            & $0.001$   &0.02   &${\small<}10^{-2}$   &${\small<}10^{-2}$\\
Tagging                               & 0.03            & ${\small<}10^{-3}$       &${\small<}10^{-3}$   &${\small<}10^{-3}$      & ${\small<}10^{-3}$   & ${\small<}10^{-3}$   &  0.04   &${\small<}10^{-2}$   & ${\small<}10^{-2}$ \\
Acceptance & 0.02      &  0.004       & 0.002    &    $0.002$       & $0.004$        & -    & -    & ${\small<}10^{-2}$    & -  \\
\hline
Models: &&& &&& &&&  \\
Default fit  &${\small<}10^{-2}$      & $ 0.003$       &${\small<}10^{-3}$    &    $0.001$       & $0.001$        & $ 0.006$   & 0.07   & 0.01   & 0.01 \\
Signal mass              &${\small<}10^{-2}$      & 0.001            &${\small<}10^{-3}$    &${\small<}10^{-3}$       & $ 0.001$     & $ {\tiny<}10^{-3}$    & 0.03   & 0.04   & 0.01 \\ 
Background mass    &${\small<}10^{-2}$      & 0.001            & $ 0.001$     & ${\small<}10^{-3} $     &${\small<}10^{-3}$      & $0.002$    & 0.06   & 0.02   & 0.02 \\ 
Resolution              & 0.02            &${\small<}10^{-3}$     & 0.001          & $ 0.001$        &${\small<}10^{-3}$     & $0.002$    & 0.04   & 0.02   & 0.01\\ 
Background time  & 0.01            & 0.001            &${\small<}10^{-3}$      & $0.001$        &${\small<}10^{-3}$      & $0.002$     & 0.01   & 0.02   & 0.02 \\ 
Background angles  & 0.02            & 0.008            & 0.002          & 0.008              & 0.009            & 0.027           &  0.06   & 0.07   & 0.03 \\
\hline					  
\bf Total  & \pSsyst & \DGsyst & \GSsyst  & \Apasqsyst & \Azesqsyst  & \ASsqsyst & \Deltperpsyst & \Deltparasyst & \DeltPSsyst \\
\end{tabular}
\end{ruledtabular}
\end{center}
\caption{Summary of systematic uncertainties assigned to the physics parameters.}
\label{tab:syst_totals}
\end{table*}

\section{Discussion}
\begin{figure*}
\begin{center}
\includegraphics[width=0.43\textwidth]{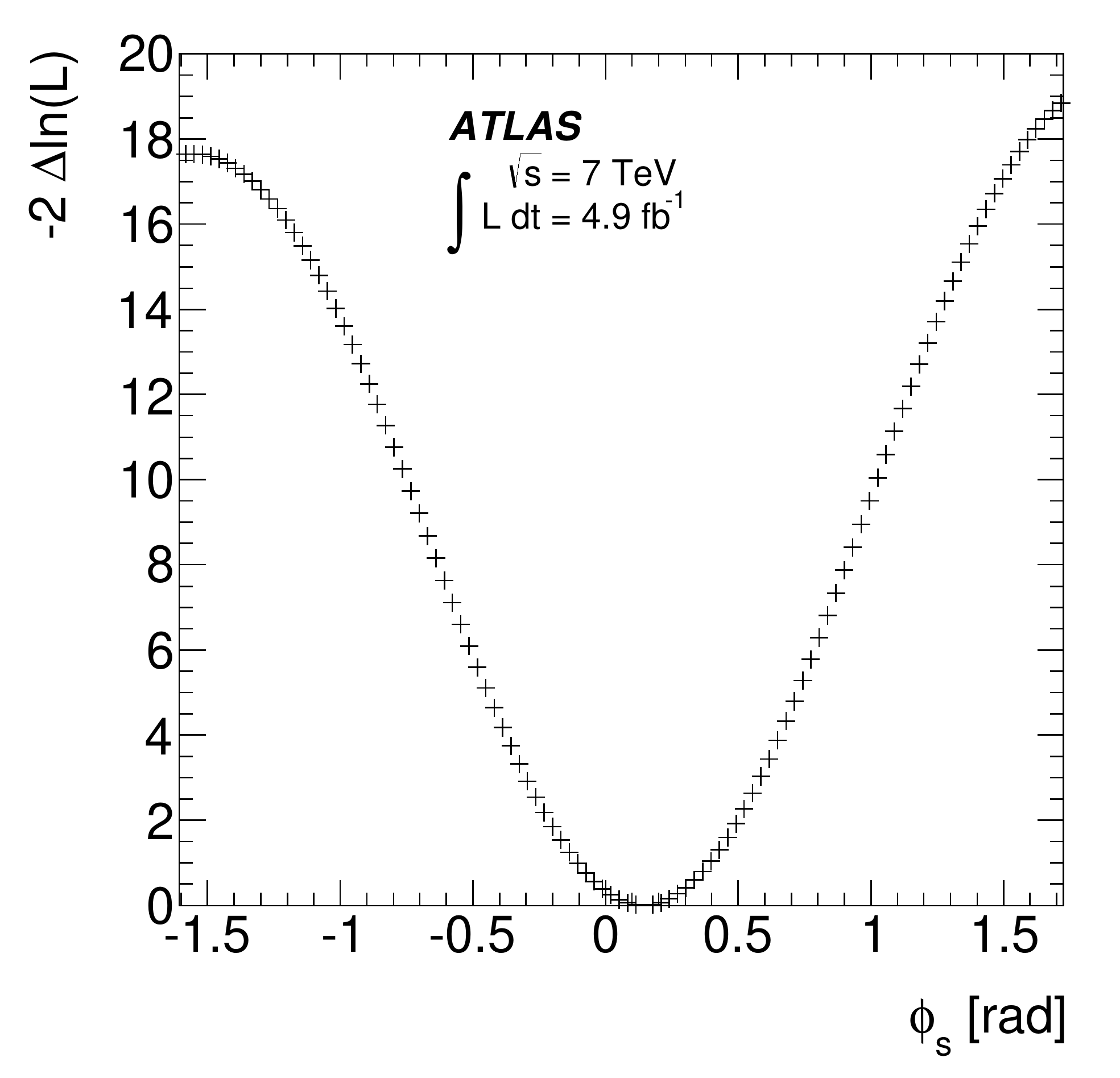}
\includegraphics[width=0.43\textwidth]{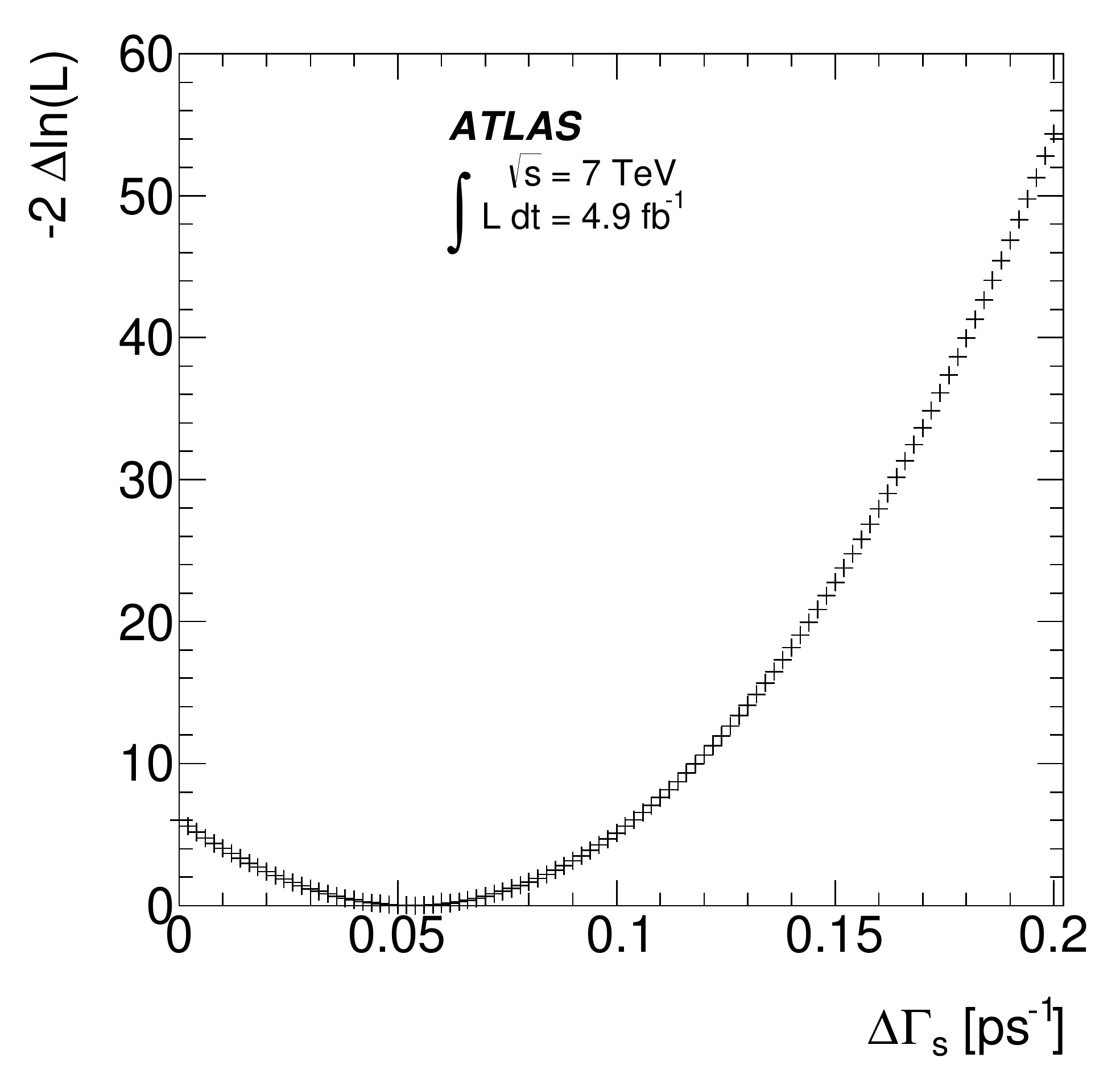}
\caption{1D likelihood scans for  \phis\ (left) and \DGs\ (right). }
\label{fig:phiSDeltaGLikelihoodScans}
\end{center}
\end{figure*}

\begin{figure*}
\begin{center}
\includegraphics[width=0.66\textwidth]{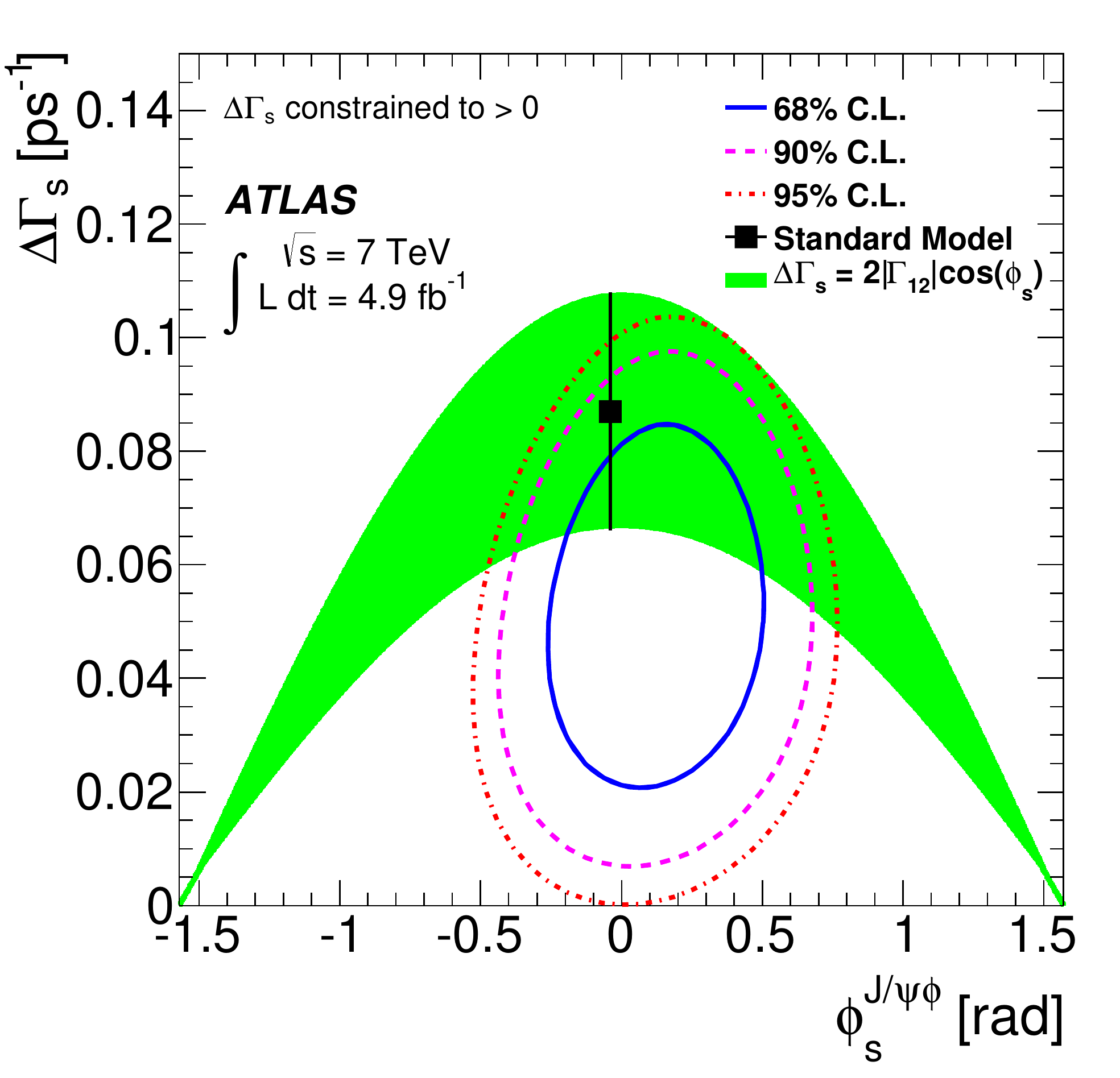}
\caption{Likelihood contours in the \phis\ -- \DGs\ plane. The blue line shows the 68\% likelihood contour, the dashed pink line shows the 90\% likelihood contour and the red dotted line shows the 95\% likelihood contour (statistical errors only). The green band is the theoretical prediction of mixing- induced $CP$ violation.}
\label{fig:Contour}
\end{center}
\end{figure*}

The PDF describing the \Bst\ decay is invariant under the following simultaneous transformations: 
\begin{equation}
\{\phis, \DGs, \delta_{\perp}, \delta_{\parallel}\}  \rightarrow \{\pi -\phis, -\DGs, \pi - \delta_{\perp}, 2\pi - \delta_{\parallel}\} \nonumber
\end{equation}
\DGs\ has been determined to be positive~\cite{Aaij:2012eq}.  Therefore there is a unique solution and only the case \DGs$~>~0$ is considered. Uncertainties on individual parameters were studied in detail in likelihood scans. Figure \ref{fig:phiSDeltaGLikelihoodScans} shows the 1D likelihood scans for \phis\ and \DGs.  Figure \ref{fig:Contour} shows the likelihood contours in the \phis\ -- \DGs\ plane.

\begin{figure*}
\begin{center}
\includegraphics[width=0.43\textwidth]{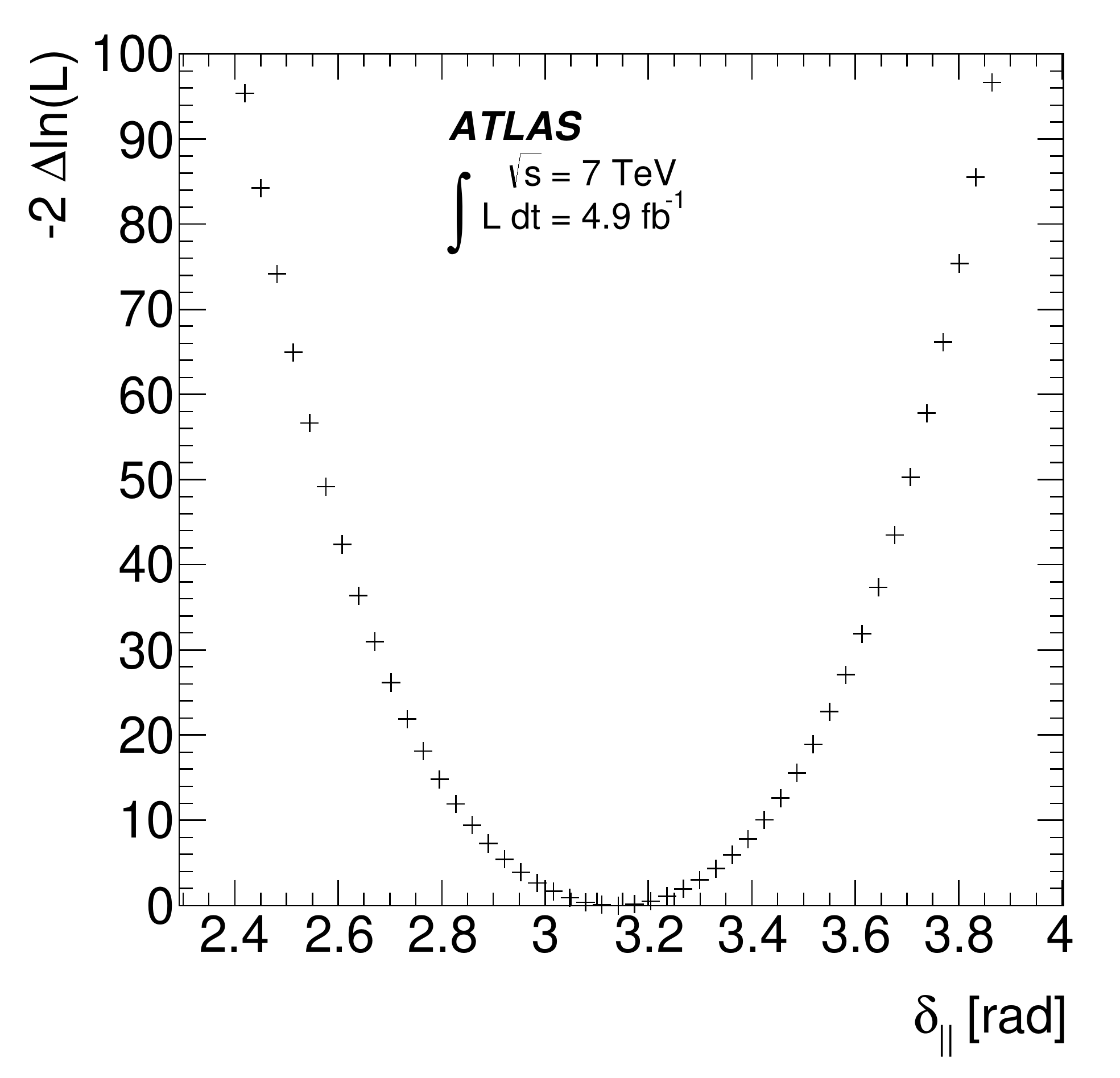}
\includegraphics[width=0.43\textwidth]{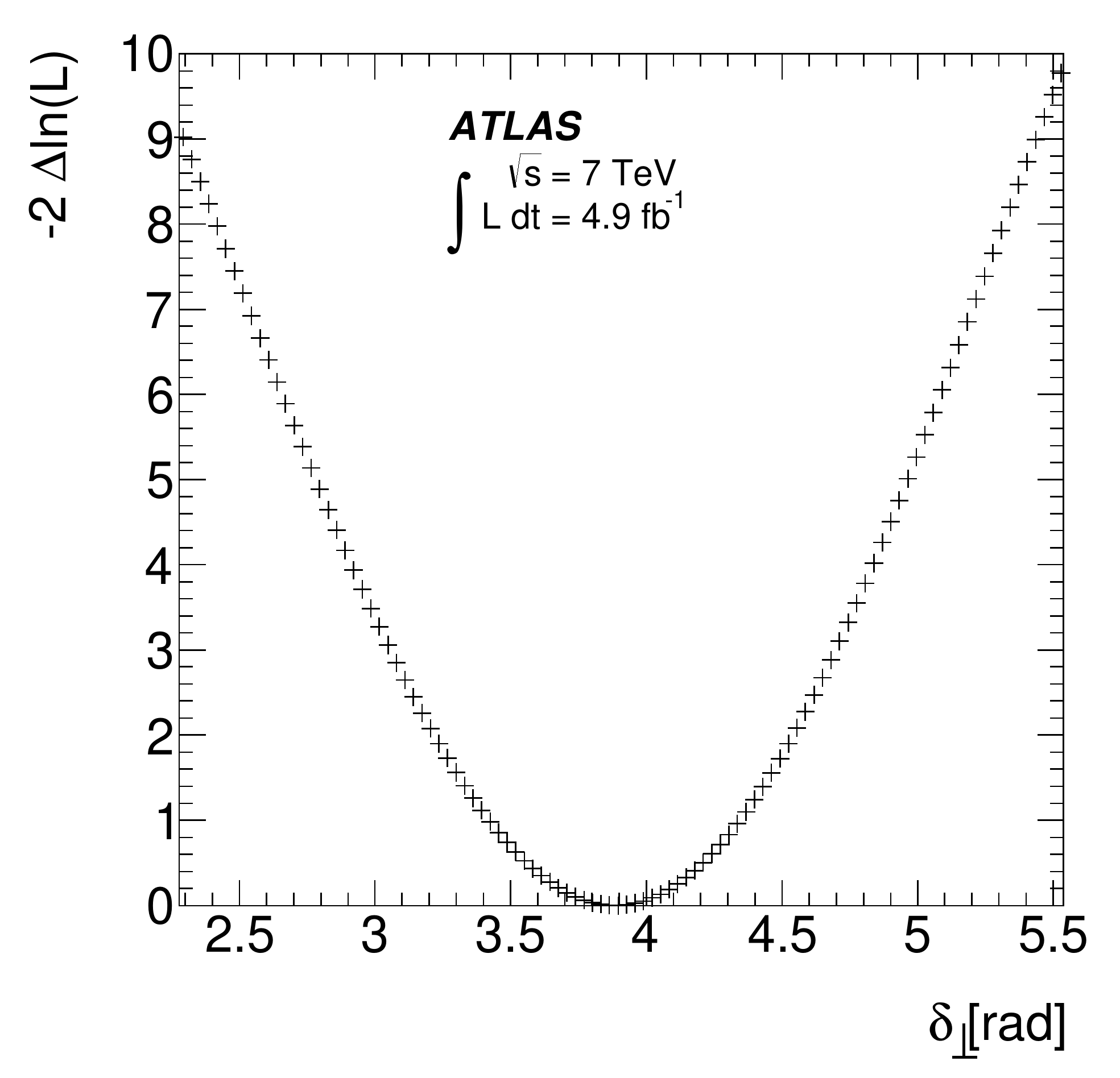}
\includegraphics[width=0.43\textwidth]{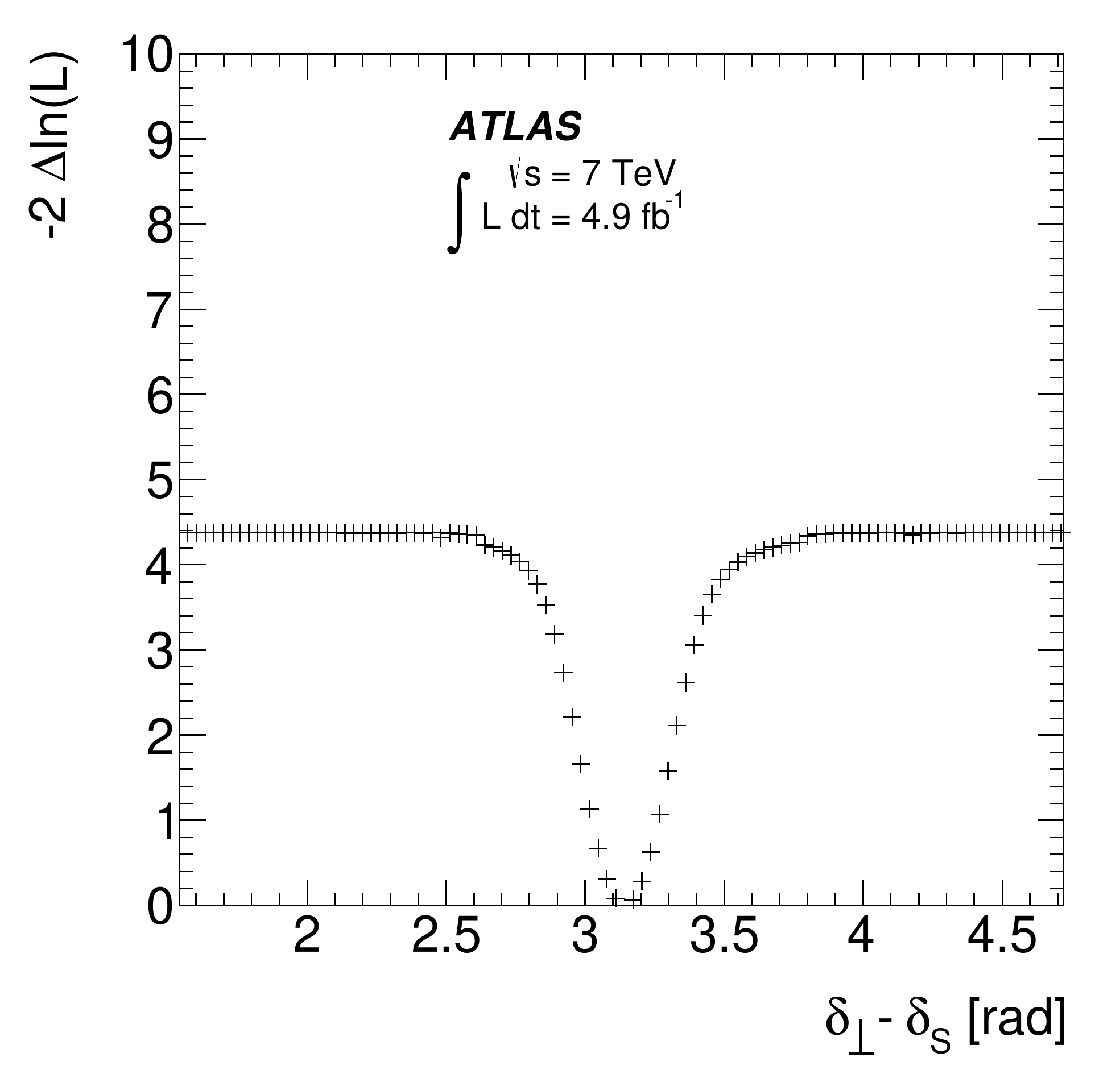}
\caption{1D likelihood scans for $\delta_{||}$ (top left), $\delta_{\perp}$ (top right) and $\delta_\perp - \delta_S$ (bottom).}
\label{fig:deltaLikelihoodScans}
\end{center}
\end{figure*}

The behaviour of the amplitudes around their fitted values is Gaussian, however the strong phases are more complicated.  Figure \ref{fig:deltaLikelihoodScans} shows the 1D likelihood scans for the three measured strong phases.  

The likelihood behaviour of $\delta_\perp$ appears Gaussian and therefore it is reasonable to quote $\delta_\perp = 3.89 \pm 0.47(stat)$ rad.  For $\delta_{\perp}-\delta_S$ the scan shows a minimum close to $\pi$; however, it is insensitive over the rest of the scan at the level of 2.1$\sigma$. Therefore the measured value of the difference $\delta_{\perp}-\delta_S$ is only given as 1$\sigma$ confidence interval \DeltPSfit~rad.   For the strong phase $\delta_{||}$ the central fit value is close to $\pi$ $(3.14 \pm 0.10)$ and the 1D likelihood scan shows normal Gaussian behaviour around this minimum.   However, the systematic pull plot based on 2400 pseudo-experiments fits reveals a double-Gaussian shape with $68\%$ of the results included in the interval $[2.92, 3.35]$~rad and so we quote the result in the form of a $68\%$  C.L. interval  $\delta_{||} \in [2.92, 3.35]$~rad (statistical only).

\newpage

\section{Conclusion}
A measurement of time-dependent $CP$ asymmetry parameters in \Bsto\ decays from a \ilumi\ data sample of $pp$ collisions collected with the ATLAS detector during the 2011 $\rts = 7\TeV$ LHC run is presented.  Several parameters describing the \Bs\ meson system are measured. These include the mean \Bs\ lifetime $1/ \Gs$, the decay width difference \DGs\ between the heavy and light mass eigenstates, and the transversity amplitudes $|A_{0}(0)|$ and $|A_{\parallel}(0)|$.  Each of these is consistent with its respective world average.  Likelihood contours in the \phis\ - \DGs\ plane are also provided. The fraction $|A_{S}(0)|^2$, the signal contribution from $\Bs \ra J/\psi K^+K^-$ and $\Bs \ra J/\psi f_{0}$ decays,  is measured to be consistent with zero, at $\ASsqfit \pm \ASsqstat~\rm{(stat.)} \pm \ASsqsyst~\rm{(syst.)}$. 

The results are:
\begin{eqnarray*}
\phis & = & \pSfit \pm \pSstat~\rm{(stat.)}\pm \pSsyst~\rm{(syst.)}~rad  \\
\DGs & = & \DGfit \pm \DGstat~\rm{(stat.)}\pm \DGsyst~\rm{(syst.)~ps}^{-1} \\
\Gs & = & \GSfit \pm \GSstat ~\rm{(stat.)}\pm \GSsyst ~\rm{(syst.)~ps}^{-1}\\
|A_{\parallel}(0)|^2 & = & \Apasqfit \pm \Apasqstat~\rm{(stat.)}\pm \Apasqsyst~\rm{(syst.)}\\
|A_{0}(0)|^2 & = & \Azesqfit \pm \Azesqstat~\rm{(stat.)}\pm \Azesqsyst~\rm{(syst.)}\\
\delta_\perp & = & \Deltperpfit \pm \Deltperpstat ~\rm{(stat.)}\pm \ \Deltperpsyst~\rm{(syst.)}~rad \\
\end{eqnarray*}
The values are consistent with those obtained in our untagged analysis \cite{ATLASBsJpsiphi} and significantly reduce the overall uncertainty on $\phis$. These results are consistent with the values predicted in the Standard Model.

\section*{Acknowledgments}
We thank CERN for the very successful operation of the LHC, as well as the
support staff from our institutions without whom ATLAS could not be
operated efficiently.

We acknowledge the support of ANPCyT, Argentina; YerPhI, Armenia; ARC,
Australia; BMWF and FWF, Austria; ANAS, Azerbaijan; SSTC, Belarus; CNPq and FAPESP,
Brazil; NSERC, NRC and CFI, Canada; CERN; CONICYT, Chile; CAS, MOST and NSFC,
China; COLCIENCIAS, Colombia; MSMT CR, MPO CR and VSC CR, Czech Republic;
DNRF, DNSRC and Lundbeck Foundation, Denmark; EPLANET, ERC and NSRF, European Union;
IN2P3-CNRS, CEA-DSM/IRFU, France; GNSF, Georgia; BMBF, DFG, HGF, MPG and AvH
Foundation, Germany; GSRT and NSRF, Greece; ISF, MINERVA, GIF, I-CORE and Benoziyo Center,
Israel; INFN, Italy; MEXT and JSPS, Japan; CNRST, Morocco; FOM and NWO,
Netherlands; BRF and RCN, Norway; MNiSW and NCN, Poland; GRICES and FCT, Portugal; MNE/IFA, Romania; MES of Russia and ROSATOM, Russian Federation; JINR; MSTD,
Serbia; MSSR, Slovakia; ARRS and MIZ\v{S}, Slovenia; DST/NRF, South Africa;
MINECO, Spain; SRC and Wallenberg Foundation, Sweden; SER, SNSF and Cantons of
Bern and Geneva, Switzerland; NSC, Taiwan; TAEK, Turkey; STFC, the Royal
Society and Leverhulme Trust, United Kingdom; DOE and NSF, United States of
America.

The crucial computing support from all WLCG partners is acknowledged
gratefully, in particular from CERN and the ATLAS Tier-1 facilities at
TRIUMF (Canada), NDGF (Denmark, Norway, Sweden), CC-IN2P3 (France),
KIT/GridKA (Germany), INFN-CNAF (Italy), NL-T1 (Netherlands), PIC (Spain),
ASGC (Taiwan), RAL (UK) and BNL (USA) and in the Tier-2 facilities
worldwide.

\bibliographystyle{apsrev4-1}

\onecolumngrid
\clearpage
\begin{flushleft}
{\Large The ATLAS Collaboration}

\bigskip

G.~Aad$^{\rm 84}$,
T.~Abajyan$^{\rm 21}$,
B.~Abbott$^{\rm 112}$,
J.~Abdallah$^{\rm 152}$,
S.~Abdel~Khalek$^{\rm 116}$,
O.~Abdinov$^{\rm 11}$,
R.~Aben$^{\rm 106}$,
B.~Abi$^{\rm 113}$,
M.~Abolins$^{\rm 89}$,
O.S.~AbouZeid$^{\rm 159}$
% ...
\bigskip
\\
$^{1}$ Department of Physics, University of Adelaide, Adelaide, Australia\\
% ...
$^{11}$ Institute of Physics, Azerbaijan Academy of Sciences, Baku, Azerbaijan\\
% ...
$^{21}$ Physikalisches Institut, University of Bonn, Bonn, Germany\\
% ...
$^{84}$ CPPM, Aix-Marseille Universit{\'e} and CNRS/IN2P3, Marseille, France\\
% ...
$^{89}$ Department of Physics and Astronomy, Michigan State University, East Lansing MI, United States of America\\
% ...
$^{106}$ Nikhef National Institute for Subatomic Physics and University of Amsterdam, Amsterdam, Netherlands\\
% ...
$^{112}$ Homer L. Dodge Department of Physics and Astronomy, University of Oklahoma, Norman OK, United States of America\\
$^{113}$ Department of Physics, Oklahoma State University, Stillwater OK, United States of America\\
% ...
$^{116}$ LAL, Universit{\'e} Paris-Sud and CNRS/IN2P3, Orsay, France\\
% ...
$^{152}$ Institute of Physics, Academia Sinica, Taipei, Taiwan\\
% ...
$^{159}$ Department of Physics, University of Toronto, Toronto ON, Canada\\
% ...
\end{flushleft}

\end{document}